\documentclass[10pt]{article}

\usepackage{amsmath}
\usepackage{amssymb}

\usepackage{graphicx}

\usepackage{cite}

\usepackage{color,soul} 

\usepackage[labelfont=bf,labelsep=period,justification=raggedright]{caption}

\makeatletter
\renewcommand{\@biblabel}[1]{\quad#1.}
\makeatother

\date{}

\begin{document}

\begin{flushleft}
{\Large
\textbf{Epigenetic landscapes explain partially reprogrammed cells and identify key reprogramming genes}
}
\\
Alex H. Lang$^{1,2,\ast}$, 
Hu Li$^{3,4,\dagger}$, 
James J. Collins$^{3,4,5,6}$
Pankaj Mehta$^{1,2,\ast}$
\\
\bf{1} Department of Physics, Boston University, Boston, MA, USA
\\
\bf{2} Center for Regenerative Medicine, Boston University, Boston, MA, USA
\\
\bf{3} Department of Biomedical Engineering, Boston University, Boston, MA, USA
\\
\bf{4} Wyss Institute for Biologically Inspired Engineering, Harvard University, Boston, MA, USA
\\
\bf{5} Howard Hughes Medical Institute, Boston, USA
\\
\bf{6} Center for BioDynamics, Boston University, Boston, MA, USA
\\
$\dagger$ Current Address: Center for Individualized Medicine, Mayo Clinic, Rochester, MN, USA
\\
$\ast$ E-mail: alexlang@bu.edu, pankajm@bu.edu
\end{flushleft}

\section*{Abstract}
A common metaphor for describing development is a rugged ``epigenetic landscape'' where cell fates are represented as attracting valleys resulting from a complex regulatory network. Here, we introduce a framework for explicitly constructing epigenetic landscapes that combines genomic data with techniques from spin-glass physics. Each cell fate is a dynamic attractor, yet cells can change fate in response to external signals. Our model suggests that partially reprogrammed cells are a natural consequence of high-dimensional landscapes and predicts that partially reprogrammed cells should be hybrids that co-express genes from multiple cell fates. We verify this prediction by reanalyzing existing datasets. Our model reproduces known reprogramming protocols and identifies candidate transcription factors for reprogramming to novel cell fates, suggesting epigenetic landscapes are a powerful paradigm for understanding cellular identity.

\section*{Author Summary}
Traditionally, standard development has been viewed as a one-way process; an organism starts as a single cell (embryonic stem cell, ESC) that divides into a multitude of mature cell types (skin cells, heart, liver, etc). But, in 2006 Takahashi and Yamanaka revolutionized this view by stochastically converting skin cells into cell types resembling ESC (called induced pluripotent stem cells, iPSC). Following this groundbreaking experiment, other reprogramming protocols have been found so now scientists can switch between a variety of cell types such as ESC, skin, liver, neurons, and heart. This has already revolutionized the understanding of biology and could change the future of medicine. A common metaphor for development is Waddington's landscape, in which an ESC is like a ball rolling down a hill which eventually ends in a valley (mature cell type). In this paper, we make this analogy more precise by developing a mathematical model of cellular development. Using data on real cell types, we can provide insight into existing reprogramming protocols and potentially predict new reprogramming protocols.

\section*{Introduction}

Understanding the molecular basis of cellular identity and differentiation is a major goal of  modern biology. This is especially true in light of the  work of Takahashi and Yamanaka demonstrating that the overexpression of just four transcription factors (TFs) is sufficient to convert somatic fibroblasts into cells resembling embryonic stem cells (ESCs), dubbed induced pluripotent stem cells (iPSCs) \cite{Takahashi2006Induction}. The idea of using a small set of TFs to reprogram cell fate has proven to be extremely versatile and reprogramming protocols now exist for  generating neurons \cite{Vierbuchen2010Direct}, cardiomyocytes \cite{Ieda2010Direct}, liver cells \cite{Sekiya2011Direct,Huang2011Induction}, neural progenitor cells (NPC) \cite{Lujan2012Direct}, and thyroid \cite{Antonica2012Generation} (see reviews \cite{Gonzalez2011Methods,Vierbuchen2012Molecular} for more details).
Despite these revolutionary experimental advances, cell fate is still poorly understood mechanistically and theoretically. Recent experiments suggest cell fates can be viewed as high-dimensional attractor states of the gene regulatory networks underlying cellular identity \cite{Huang2005Cell}. In particular, cell fates are characterized by  a robust gene expression  and epigenetic state resulting from the complex interplay of transcriptional regulation, chromatin regulators, non-coding and microRNAs, and signal transduction pathways. 


These experiments  have renewed interest in the idea of an  `epigenetic landscape'  that underlies cellular identity \cite{Waddington1957The-Strategy,Kauffman1993The-Origins,Enver2009Stem,Zhou2011Understanding,Ferrell2012Bistability}. The landscape picture requires several key features to be consistent with experimental observations  (see Figure 1). All cell fates must be robust attractors, yet allow cells to change fate through rare stochastic transitions \cite{Gonzalez2011Methods,Buganim2012Single-Cell} as in cellular reprogramming experiments (Figure 1A). A common result of reprogramming is not the desired cell fate, but partially reprogrammed cells \cite{Sridharan2009Role,Mikkelsen2008Dissecting}. These results suggest that the landscape is rugged and may contain additional spurious attractors corresponding to  cell fates that do not naturally occur \emph{in vivo}. In addition, environmental and external signals can control cell fates. Some environments stabilize particular cell fates (Figure 1B). A dramatic example of this is a protocol for reprogramming  to neural progenitor cells (NPCs) that is identical to Yamanaka's protocol for reprogramming to ESC except for the culturing media \cite{Kim2011Direct}. Other external signals deterministically switch cell fates, as occurs in normal development (Figure 1C) \cite{Davidson2006The-Regulatory}. Together, these imply the landscape is a dynamic entity that depends on environmental signals.

The recent experimental progress has inspired several different theoretical approaches to understand the epigenetic landscape and the underlying gene regulatory networks governing cell fates. One focus has been on explicit construction of landscapes for specific cell fate decisions such as the erythroid vs myeloid choice  in hemopoietic development \cite{Huang2007Bifurcation}, pancreatic cell fates \cite{Zhou2011Predicting}, or \emph{C. elegans} vulva development \cite{Corson2012Geometry}. Other network based approaches use experimental data to constrain the possible networks \cite{Henry2013Network,Zagorski2013Edge}. A second area of work is based on understanding the underlying gene regulatory network \cite{MacArthur2009Systems,Lu2009Systems-level}. A recent paper \cite{Banerji2013Cellular} attempts to combine the network and landscape picture by using the network entropy to define a landscape.  On a more abstract level, there has been a renewed interest in understanding Waddington's landscape mathematically using ideas from dynamical systems and nonequilibrium statistical mechanics \cite{Ferrell2012Bistability, Wang2011Quantifying-Waddington}.  Most of these models focus on \emph{in vivo} developmental decisions and hence consider the dynamics of a few genes or proteins.

Here, we present a new modeling framework to construct a global (i.e. all cell fates and all TFs) epigenetic landscape that combines techniques from spin glass physics with whole genome expression profiles. We were inspired by the successful application of  spin glasses to model neural networks \cite{Hopfield1982Neural,Amit1985Spin-glass,Kanter1987Associative,Amit1992Modeling} and protein folding landscapes \cite{Bryngelson1995Funnels}. Here, we construct an epigenetic landscape model for cellular identity with 63 stable cell fates and 1337 TFs using cell-fate specific, mouse microarray gene expression data.  Each cell fate is a robust attractor, yet cells can deterministically switch fates in response to external signals. Our model provides a unified framework to discuss differentiation and reprogramming. It also naturally explains the existence of partially reprogrammed cell fates  as `spurious' attractors resulting from the high dimensionality of the landscape. Our model predicts, and we verify,  that partially reprogrammed cells are hybrids that co-express TFs of multiple naturally occurring cell fates.  Finally, our model reproduces known reprogramming protocols to iPSCs, heart, liver, NPC, and thyroid, and has the potential for designing reprogramming protocols to novel cell fates. Taken together, these results suggest that epigenetic landscapes represent a powerful framework for understanding the molecular circuitry and dynamics that gives rise to cell fate.

The organization of the paper is as follows. First, we explain the motivation for using an attractor neural network to model the epigenetic landscape. Second, we define the state space for the model and the actual biological data used to construct the state space. Third, we give an overview of our landscape model (with details given in Table 1 and Materials and Methods: Landscape Model). Next, we show that our mathematical model captures the essential experimental features of cellular identity. We then show that our model naturally explains the existence of partially reprogrammed cells and makes predictions about their gene expression profiles. We verify this by reanalyzing experimental data. Finally, we show that our model can identify key reprogramming genes in existing reprogramming protocols, suggesting it can be used to identify candidate TF for reprogramming to novel cell fates. We conclude by discussing the implications of our mathematical model for understanding cellular identity and reprogramming.

\section*{Results}

\subsection*{Motivation from attractor neural networks}
The Takahashi and Yamanaka reprogramming experiments \cite{Takahashi2006Induction} are reminiscent of content-addressable memory and attractor neural networks. First, let us introduce a content-addressable memory with a paraphrasing of the original Hopfield paper. A content-addressable memory allows one to retrieve a full memory based on sufficient partial information. For example, suppose the complete stored memory is ``John J. Hopfield, Neural networks and physical systems with emergent collective computational abilities (1982).''  A content-addressable memory is capable of retrieving the full memory based on partial, incomplete input. Therefore, the details  ``Hopfield,'' ``Neural networks,'' and ``1982'' could be enough to recall the full memory.

In the Yamanaka reprogramming protocol, overexpressing only four TFs is enough for a fibroblast to ``recall'' the global TF expression of an ESC. A content-addressable memory is naturally represented as a basin of attraction in a dynamical system, with partial recall corresponding to entering the basin of attraction and full recall corresponding to reaching the minimum of the basin. Hopfield attractor neural networks \cite{Hopfield1982Neural,Amit1985Spin-glass,Amit1992Modeling} are a general method to take an input set of vectors (``memories'') and explicitly construct a unique, global, landscape such that each input vector is a global minimum and has a basin of attraction. In what follows, we will exploit the analogy between associative memory in attractor neural networks and cellular reprogramming to explicitly construct the epigenetic landscape underlying cellular identity.

\subsection*{The epigenetic landscape}
Our goal is to model the global epigenetic landscape involving all cell fates by using genome wide data. Currently, microarrays are the only technology with genome wide data for a multitude of cell fates (although RNA-seq and other technologies will likely be useful in the future). Specifically, we compiled a dataset of 601 mouse whole genome microarrays (details in Materials and Methods: Data Analysis) resulting in the gene expression for $N=1337$ transcription factors for $p=63$ cell fates. We restricted our considerations to TFs due to their importance in cellular reprogramming and differentiation. However, our model can be easily generalized to include other important genes. To robustly compare microarrays from multiple platforms, we converted the raw expression data into a rank ordered list. We assumed that gene expression is log-normal distributed (the minimal-assumption model for positive-definite random numbers such as gene expression) and assigned a z-score to each TF.  The final output of this procedure is that it assigns each TF in every cell fate a z-score gene expression.

This continuous gene expression could be used to construct our epigenetic landscapes. However, for mathematical convenience, we discretize the continuous gene expression data into high expression  ($+1$ for z-score $>=0$) and low expression ($-1$ for z-score $<0$). See Text S1 for an extended discussion on continuous vs discrete TF expression in attractor neural networks.

This discretization process is biologically plausible. Cellular identity and differentiation are largely controlled by epigenetics, especially histone modifications (HMs) \cite{Jenuwein2001Translating} (Figure 2A).  Epigenetics primarily controls the accessibility of DNA and depending on the HM, the DNA can be stabilized in an open or closed configuration. Using global HM data  \cite{Mikkelsen2007Genome-wide,Meissner2008Genome-scale} and comparing it to microarray data, we created a conditional probability distribution of having a HM  given a TF expression level (Figure 2B). We find that between a z-score of $-0.5$ to $0.5$ there is a sharp threshold which distinguishes genes with the activating modification of histone 3 tri-methylation at lysine 4 (K4) from genes with the inactivating modification of histone 3 tri-methylation at lysine 27 (K27) and poised/bivalent genes (both K4 and K27). This provides a potential biological justification to our discretization. In summary, we take the continuous gene expression and binarize (Figure 2C). These binary (i.e. on/off) TF data are the only biological input into our model. 

In order to precisely describe the landscape results, we need to define the correct way to measure distances. One possible measure is the overlap (aka dot product or magnetization), defined for cell fate $\mu$ as:
\begin{equation}
m^\mu=\frac{1}{N}\sum_{i=1}^N \xi_i^\mu S_i
\end{equation}
where $S_i$ is an arbitrary expression state and $\xi_i^\mu$ is the gene expression in the natural cell fate $\mu$. The overlap between cell fate $\mu$ and state $S_i$ for exactly correlated, anti-correlated, or uncorrelated states is $1$, $-1$, or $0$ respectively. 

Cell fates from similar lineages (ex. blood) often have similar gene expression patterns.  For example,  B cells and T cells have a  77\% overlap in their gene expression profiles. Such large correlations between cell fates makes the overlap, $\textbf{m}$, a poor distance measure. In order to measure distances between highly correlated vectors, it is helpful to define the ``projection'' $a^\mu$  of a gene  expression state $S_i$ on a cell fate $\mu$  by
\begin{equation}
a^\mu=\sum_{\nu=1}^p (A^{-1})^{\mu\nu}m^\nu
\end{equation}
where $A^{-1}$ is the inverse correlation matrix and $m^\nu$ is the overlap on cell fate $\nu$ and is given by
\begin{equation}
A^{\mu\nu}=\frac{1}{N}\sum_{i=1}^N \xi_i^\mu \xi_i^\nu
\end{equation}
The projection $a^\mu$ measures the orthogonal projection of a state $S_i$ onto the subspace spanned by naturally occurring cell fates, $\xi$ (see Figure 2D and Text S1), and a perfect projection onto state $\mu$ is given by $a^\mu=1$.  In contrast with the overlap,  B cells have zero projection on T cells, and vice versa. 

Our landscape  assigns an ``energy'' to every global expression state. We emphasize that this energy does not correspond to physical energy consumption of ATP; instead it is an abstract energy that corresponds to stability and  developmental potential of cell fates. The complete landscape $H$ can be thought of as arising from  four terms with a simple interpretation (see Figure 1):
\begin{eqnarray}
H &=&  H_{basin} + H_{bias} + H_{culture} + H_{switch} \label{landscape}
\end{eqnarray}
The first term, $H_{basin}$, ensures that observed cell fates are valleys in our landscape (Figure 1A). The second term, $H_{bias}$,  describes biasing of specific TFs by experimentalists (not shown in Figure 1). The third term, $H_{culture}$, increases the radius and depth of  cell fates that are favored by the environment or culturing conditions (Figure 1B). Finally, in the presence of an external signal that gives rise to differentiation (ex. growth factors associated with differentiation), the fourth term, $H_{switch}$, opens a low energy path between the initial and final cell fates (Figure 1C). We give a complete mathematical description of the model in the Materials and Methods: Landscape Model and a summary in Table 1.

\subsection*{Cell fates are dynamic attractors that are responsive to signals}

We performed self-consistency checks for our model using two \emph{in silico} experiments (see details in Materials and Methods: Simulations). To verify that naturally occurring cell fates are dynamic attractors, we randomly perturbed the gene expression profile of cells from the ESC state and then tracked the gene expression over time. Real biology has many potential sources of noise, and the asynchronous dynamics introduced above will likely underestimate the noise. To show that our model is still robust to other large sources of noise, in our simulations we also add in periodic bursts of noise by flipping a fixed percentage of TF states (2\%) to mimic the observation that cellular divisions produce HM errors \cite{Ben-David2011Large-Scale}. Figure 2E shows the projection of the TF state on the ESC state as a function of time.  For a large number of starting conditions, after an initial transient, the system relaxes back to the ESC state (red bracket), explicitly demonstrating the existence of a large basin of attraction \cite{Huang2005Cell}. This is true even when we  break detailed balance by making the interaction matrix asymmetric by randomly deleting  20\% of interactions (Figure 2E Diluted).

Our model can also deterministically switch between cell fates in response to differentiation signals. For example, the common myeloid progenitor (CMP) is a blood cell fate that \emph{in vivo} can differentiate into either granulo-monocytic progenitors (GMP) or megakaryocyte-erythroid progenitors (MEP). In Figure 2F, we show \emph{in silico} validation where we start the system in the CMP state and show the trajectories after applying either the GMP (signal 1, blue) or MEP (signal 2, red) differentiation signal, resulting in branching to two distinct cell fates.

\subsection*{Partially reprogrammed cells as ``spurious'' attractors}

When performing a reprogramming experiment, besides the initial cell fate and the end goal cell fate, experimentalists often produce ``novel cell fates'', dubbed partially reprogrammed cells \cite{Sridharan2009Role,Mikkelsen2008Dissecting}. These partially reprogrammed cells have the characteristics of a stable cell fate (i.e. they can be passaged indefinitely in culture), but may express a mix of key markers for multiple cell fates and have a global gene expression that does not match any \emph{in vivo} cell fate \cite{Mikkelsen2008Dissecting}. 

While the existence of partially reprogrammed cells was surprising to experimentalists, they have a natural interpretation in our model. One of the most generic properties of all attractor neural networks is that in addition to the desired attractors, $\xi_i^\mu$, the non-linearity of the dynamical process and  topology of high-dimensional (in our case $N=1337$) vector spaces induces additional attractors, which are termed spurious attractors \cite{Amit1992Modeling}. In our model, since the natural cell fates are the input vectors, these spurious attractors can be interpreted as potential cell fates that do not occur \emph{in vivo}. These spurious attractors are predicted to be low-dimensional combinations, or hybrids (see Materials and Methods: Spurious Attractors and Text S1 for details) that should also be stable attractors but with smaller basins of attraction.

A priori, there are several valid hypotheses for the relationship between partially reprogrammed cells and natural cell fates. In the original experiments \cite{Sridharan2009Role,Mikkelsen2008Dissecting}, it was expected that partially reprogrammed cells should be a hybrid of the starting and goal cell fate only  (i.e. have a significant projection only on the starting or ending cell fate). Another hypothesis was that in a high-dimensional landscape, randomly chosen vectors should be orthogonal (Figure S2) (i.e. have a projection of $a\approx0$ with all cell fates). However, our model predicts that partially reprogrammed cells should be low-dimensional hybrids of existing cell fates, but that they do not necessarily have to be a combination of the starting and goal cell fate. Mathematically, we predict that partially reprogrammed cells should only have a projection $|a|>0.106$ (2 std above 0, see Figure S2) for a small number of natural cell fates. Reanalyzing existing genome-wide datasets on partially reprogrammed cells (Table 2) validates the prediction of our model that partially reprogrammed cells are low-dimensional hybrids of existing cell fates. This qualitative agreement between the predicted spurious attractors and the partially reprogrammed states is independent of details of our landscape function. Importantly, such hybrid states are a generic property of \emph{all} attractor-based landscape models and hence represents an important criteria for judging whether attractor-based models are suitable for describing epigenetic landscapes. 

\subsection*{Identifying transcription factors for cellular reprogramming}

Our landscape model provides a quantitative method to identify ``predictive'' TFs for a given cell fate. These predictive TFs can be used as markers of a cell fate and are potential candidates for reprogramming protocols.  We expect reprogramming TFs to be a subset of all predictive TFs but not all predictive TFs will lead to successful reprogramming. For example, cell-specific downstream targets of reprogramming TFs are likely to also be highly predictive for a cell type but may not lead to successful reprogramming.

Most reprogramming experiments follow an experimental protocol similar to the one outlined by Takahashi and Yamanaka in their seminal paper \cite{Takahashi2006Induction, Gonzalez2011Methods}. Initially the starting cells (usually mouse embryonic fibroblasts, MEFs) are infected with viruses  containing  all the TFs of interest. The original Yamanaka experiment over-expressed 24 TFs \cite{Takahashi2006Induction}, while more recent experiments usually start with about 10 TFs \cite{Vierbuchen2010Direct,Ieda2010Direct,Sekiya2011Direct,Huang2011Induction,Lujan2012Direct}. Several days after infection, the cells are switched to culturing conditions that support the desired final cell fate. If an experiment is successful, cells resembling the desired cell fate will appear after a few weeks. This original list is then pruned to identify a ``minimal'' (essential) set of TFs that still allows for successful reprogramming. In many cases, the viruses are excised \cite{Sommer2010Excision} to confirm that the the reprogramming does not depend on viral expression.  Furthermore, recent experiments indicate that the same TFs can be used to reprogram to a desired cell fate  from multiple initial cell fates \cite{Buganim2012Single-Cell}.  These experiments suggest that reprogramming TFs should be based  on  final, not initial, cell fate.

Intuitively, reprogramming candidates should be both highly expressed and highly ``predictive'' of the desired cell fate. 
The TF z-score naturally defines high and low TF expression levels. Within our landscape, the ``predictivity'' $\eta_i^\mu$ of the $i^{th}$ TF for a given cell fate $\mu$,  is measured by its contribution to the potential energy of that cell fate, and is mathematically defined as:
\begin{equation}
\eta_i^\mu = \displaystyle\sum_{\nu=1}^p \left( A^{-1}\right)^{\mu\nu} \xi_i^\nu
\end{equation}
where $A^{-1}$ is the cell fate correlation matrix and $ \xi_i^\nu$ is the expression of TF $i$ in cell fate $\nu$. We note that the projection and predictivity are directly related as can be seen by
\begin{equation}
a^\mu = \displaystyle\sum_{i=1}^N \eta_i^\mu S_i
\end{equation}
where $ \eta_i^\mu$ is the predictivity of TF $i$ in cell fate $\mu$ and $S_i$ is an arbitrary gene expression state.

For a desired target cell fate, TFs that are high (low) in both predictivity and expression in that cell fate are candidates for over expression (knock out) in reprogramming (see Figure 3A). For a simple, single measure of reprogramming efficacy of a TF, the predictivity and expression can be multiplied together to give a ``reprogramming score'', where the top (bottom) rank order TFs are the best candidates for over expression (knock out). Figure 3 shows the expression and predictivity for TFs in a variety of cell fates. In Figure 3B, we have explicitly labeled the TFs  used in the original Yamanaka protocol for reprogramming to ESC. Consistent with our model, these TFs are both predictive and highly expressed. Figure 3C shows TFs that have been successfully used in any reprogramming protocol to ESCs \cite{Gonzalez2011Methods} as well as the pluripotency genes (involved in maintaining stem cell fate)  \emph{Zfp42} (\emph{Rex1}) \cite{Masui2008Rex1/Zfp42} and \emph{Nr0b1} (\emph{Dax1}) \cite{Khalfallah2009Dax-1}.  Once again these genes are highly predictive for ESCs. As a further check on the biological validity of our predictions,  we analyzed the GO Annotation of our top 50 candidates  for ESC reprogramming (Table S1). Within these top TFs, 12 have successfully been used in reprogramming, 7 are known pluripotency TFs, 16 are involving in cell differentiation, while 15 have no known function and are intriguing reprogramming candidates. Taken together this suggest that we are capturing the essential biology despite minimal biological data for input.

While ESC have been studied in the most detail, recent experiments have reprogrammed (aka direct conversion) to other cell fates such as cardiomyocytes \cite{Ieda2010Direct} (Figure 3D), liver \cite{Sekiya2011Direct,Huang2011Induction} (Figure 3E),  and thyroid \cite{Antonica2012Generation}  (Figure 3F). Once again we have explicitly labeled the TFs that have been successfully used for direct conversion. Notice that all of these TFs (except \emph{Mef2c}) are highly predictive and highly expressed. Note that  \emph{p19Arf} \cite{Huang2011Induction} used in the direct conversion to liver was not differentially expressed  in our microarrays and therefore was not included in our model. 

We also examined TFs used in direct conversion to neural lineages. As discussed in \cite{Vierbuchen2010Direct}, these TFs were chosen because they were known to be important in either neurons or  neural progenitor cells (NPC).  Figure 3F and 3G show the expression and predictivity of TFs for neural progenitor cells (NPC) \cite{Lujan2012Direct} (Figure 3G), and neurons \cite{Vierbuchen2010Direct} respectively.  Induced NPC were made using a four TF cocktail consisting of \emph{Pou3f2} (\emph{Brn2}), \emph{Sox2}, and \emph{Foxg1} \cite{Lujan2012Direct}.  Our analysis shows that the first two of these TFs are predictive for NPCs while \emph{Foxg1} is predictive for neural stem cells (NSC) (see Figure S3).  Induced neurons (iN) can be made using the TFs  \emph{Myt1l}, \emph{Pou3f2}, and \emph{Ascl1} \cite{Vierbuchen2010Direct}. Consistent with their experimental design, we find that  \emph{Myt1l} is highly predictive for mature neurons,  while the remaining  TFs (\emph{Pou3f2}, \emph{Ascl1}) are predictive for NPCs.  

While it is not possible to perform statistical tests to test our examples due to the scarcity of reprogramming protocols, we performed a simple numerical exercise to gauge the predictive power of our model. The four Yamanaka factors are all in the top 50 when ranked by their Òreprogramming scoreÓ for ESCs (where the reprogramming score of a TF is defined as the product of the expression and predictivity scores of a TF). We randomly permuted TF labels and asked how often all four Yamanaka factors remained in the top 50. For a million independent permutations, this occurred only once, confirming that our model is capturing many essential aspects of cellular reprogramming.

Our results suggest that epigenetic landscapes may be useful for rationally-designing reprogramming protocols to novel cell fates. To this end, we have used our model to identify candidate TFs for reprogramming, see File S5 for the  top 50 candidates for overexpression for all cell fates and File S6 for top 50 candidates for knockouts for all cell fates.

\section*{Discussion}

A common biological metaphor used to describe development and cellular reprogramming is a rugged ``epigenetic landscape'' which emerges from  a complex gene regulatory network, with cell fates corresponding to attracting valleys in the landscape. Despite decades of biological innovation, the large number of genes and their complex interactions has prevented the quantitative modeling of a global epigenetic landscape. To meet this challenge, we have developed a new quantitative framework of cellular identity to directly model the global, high-dimensional epigenetic landscape. Using whole genome expression data, we constructed an epigenetic landscape based on techniques from spin glass physics and neural networks. Our landscape only depends on the experimentally determined gene expression of natural cell fates. Yet, it explains the existence of spurious cell fates (known as partially-reprogrammed cells) and can reproduce known reprogramming protocols to embryonic stem cells, heart, liver, thyroid, neural progenitor cells, and neurons. More importantly, our model can be used to identify candidate transcription factors for reprogramming to novel cell fates. 

An interesting question is if spurious attractors are ubiquitous throughout the landscape, why does standard development not produce partially reprogrammed cells? The key is the difference in the dynamics. In cellular reprogramming, the starting cell fate is forced to express a small number of TF and this leads to a stochastic conversion to the desired cell fate (Figure 1A). During this stochastic exploration of the landscape, there is only a weak bias towards the final state, so it is easy for the cells to get trapped in a metastable state. However, during standard development, the external signals actively reshape the landscape and open up low energy valleys between cell fates (Figure 1C). This strong bias towards the final cell state results in a deterministic switch during which the spurious attractors are only a small road bump on the path to the final cell state. Therefore, it is not a surprise that partially reprogrammed cells are only found during cellular reprogramming and not during standard development.

Epigenetic landscapes can also be used to identify important, or predictive, TFs for cell fates. The predictivity of a TF for a cell fate generalizes the idea of specificity. A TF is specific to a cell fate if it is expressed only on in a small subset of cell fates. In contrast with specificity, predictivity weighs the global correlations amongst cell fates when assessing the importance of a TF for a cell fate. Thus, the predictivity not only picks out important specific TFs, but also TFs that are lineage markers. For example, \emph{Brachyury} (\emph{T}) \cite{Yamaguchi1999T-Brachyury} is a general marker of mesodermal lineages. Since it is highly expressed in large a number of cell fates, it is not specific to any given cell fate. However, it is predictive because its expression is a strong indicator that a given cell fate is a mesodermal lineage. 

The concept of predictivity also yields new insights into the Yamanaka protocol. When the Yamanaka factors were first published, two of the four TFs, \emph{Pou5f1} (\emph{Oct4}) and \emph{Sox2} were known to be important for ESCs. In contrast, the role of the other two TFs,  \emph{Klf4} and \emph{Myc},  was not well understood \cite{Jaenisch2012Nuclear}. It was quickly shown that \emph{Myc} was was not essential to reprogramming (\emph{Oct4}, \emph{Sox2}, and \emph{Klf4} can reprogram alone), but nonetheless enhanced the efficacy of reprogramming \cite{Wernig2008c-Myc}. The importance of \emph{Klf4} was surprising given that it is neither highly expressed nor specific for ESC. However, \emph{Klf4} is highly predictive of ESC (Table S2). For this reason, our model actually explains why \emph{Klf4} is a prime candidate for reprogramming to ESCs. 

We make several experimentally verifiable predictions. First, our model predicts the partially reprogrammed cells should be hybrids of existing natural cell fates. As more partially reprogrammed cells are studied, if they are found to either have high projection on only one cell fate ($a^\mu\approx1$ for one $\mu$) or no projections on any cell fates ($a^\mu\approx0$ for all $\mu$), this would call into question whether partially reprogrammed cells are truly the spurious attractors of an attractor neural network. Second, our model can be used to identify important, or predictive, TFs for cell fates. TFs with large positive (negative) predictivity should be positive (negative) markers for a cell fate. Additionally, for cellular reprogramming we predict that TFs with large positive (negative) predictivity and expression could be over expressed (knocked out) to reprogram to a desired cell fate. Therefore, our model has several predictions that can be tested against future experimental progress in the field.

Our model has several limitations. First, a generic limitation for any method relying on microarrays to define gene expression is that one cannot distinguish between direct, causal, interactions and indirect, correlative, interactions. Therefore, predictivity can establish the importance of a gene, but further experiments are needed to determine if the predictive gene is the controller of the cell type or just a passive indicator of a cell type. Second, it fails to accurately capture the dynamics of reprogramming. Simulations of reprogramming with known protocols, such as the Yamanaka protocol, lead to rates of reprogramming that are comparable to the rates from a reprogramming simulation with a randomly selected protocol. This is likely due to the fact that cell fates are extremely stable and hence reprogramming is extremely rare. Third, our model does not directly explain the importance of the non-specific transcription factor \emph{Myc}. Many protocols use \emph{Myc} \cite{Gonzalez2011Methods}, but it can be replaced (with no deleterious effect) by short hairpin RNAs (shRNAs) \cite{Onder2012Chromatin-modifying}, or  dropped completely from protocols at the expense of speed and less efficient reprogramming \cite{Wernig2008c-Myc}. This suggests that \emph{Myc} may have an alternative role and instead of being a biasing field, $B_i$, it may instead raise the effective noise of the system (i.e. decrease $\beta$). Another limitation is that based on the currently available experimental data, our landscape construction cannot definitively be distinguished from alternative constructions. For example, the interaction network could be constructed by such that it does not weigh each cell fate equally (as is currently done). This would have the effect of changing the relative stability of cell fates. Therefore, in the absence of more experimental data, our landscape and a weighted landscape cannot be distinguished.

A popular approach to inferring landscapes from biology data are ``Maximum Entropy'' models. This method has been used to model firing neurons \cite{Schneidman2006Weak}, protein configurations \cite{Bialek2007Rediscovering,Cocco2013From}, and antibody diversity \cite{Mora2010Maximum}. The Maximum Entropy approach takes as input large samples of biological data and a set of constraints and outputs a landscape that maximizes the entropy. While Maximum Entropy models can be used to infer landscapes with basins of attraction \cite{Tkacik2014Searching}, it can quickly become a computationally challenging problem. Our approach differs from Maximum Entropy models in the following way. Since our goal is to model a landscape with basins of attractions, we make the ansatz that the landscape can be described by a Hopfield neural network. Then we insert real biological data, $\xi$, to construct the landscape exactly. Our method requires no computational inference of parameters.

There are several natural extensions of the model discussed in this paper. The landscape could be constructed with additional biological input such as other genes, microRNAs, or histone modification data. This opens up possibilities of improving upon the high reprogramming rates achieved by overexpressing microRNAs \cite{Yoo2011MicroRNA-mediated} or synthetic mRNAs \cite{Warren2010Highly}. Another attractive element of the framework presented here is that it allows for a quantitative analysis of whole genome-wide expression states (see Table 2). This is likely to yield a more accurate classification of reprogrammed cells. Finally, directed differentiation protocols \cite{Longmire2012Efficient} attempt to mimic standard development in vitro and have proven to have high efficiency and fidelity. Future work will try to use our landscape to predict the necessary signaling factors for rationally designing more efficient directed differentiation protocols. Overall, epigenetic landscapes provide a unifying framework for cell identity, reprogramming, and directed differentiation, and our results suggest these landscapes can provide crucial insight into the molecular circuitry and dynamics that gives rise to cell fate.

\section*{Materials and Methods}

\subsection*{Data Analysis}
Here we present the details of the dataset. All data used in this paper are available in the online Supplementary Information and is organized as follows:
\begin{itemize}
\item File S1: Microarray Sources.  List of all microarrays used in this paper.
\item File S2: TF Z-Score.  The z-score gene expression for each TF of natural cell fates in this paper. This data is post RMA normalization and averaging over multiple replicates for each natural cell fate.
\item File S3: TF Predictivity.  The predictivity for each TF and cell type in this paper.
\item File S4: Partially Reprogrammed Cells Z-Score.  The z-score gene expression for each TF of partially reprogrammed cell fate. This data is post RMA normalization and averaging over multiple replicates for each partially reprogrammed cell fate.
\item File S5: Overexpression Candidates.  Top overexpression candidates to reprogram to various cell fates.
\item File S6: Knock-Out Candidates.  Top knock-out candidates to reprogram to various cell fates.
\end{itemize}
An older version of this manuscript, Arxiv v3 \cite{Lang2012Epigenetic}, has additional microarrays available that are unused in this version of the text. All microarrays used in this paper were taken from the public databases ArrayExpress (www.ebi.ac.uk/arrayexpress) or GEO (www.ncbi.nlm.nih.gov/geo). See File S1: Microarray Sources for details on where to obtain raw, pre-normalized and pre-averaged data.

There are two datasets, the natural cell fates and the partially reprogrammed cells. For the natural cell fates, we only used the Affymetrix GeneChip Mouse Gene 1.0 ST platform due to the large number of available microarrays on ArrayExpress (www.ebi.ac.uk/arrayexpress) and the better technical design of the platform (1.0 ST has probe matches throughout a gene in contrast to just the 3' UTR in Affymetrix GeneChip Mouse Genome 430 2.0). There is limited data on partially reprogrammed cells so we used microarrays from Affymetrix GeneChip Mouse Genome 430 2.0.

The raw microarray data was converted to an expression level as follows. Microarray probe-to-gene map was created with Bioconductor 2.10. All raw microarray files were initially processed by robust mean averaging (RMA) in MATLAB, and genes with multiple microarray probes were averaged. We did additional processing of this output for two reasons. First, we need to compare microarrays from multiple platforms, but the standard RMA output can vary significantly from platform to platform. Second, since gene expression is a set of positive definite numbers, the minimal assumption model of gene expression is a log-normal distribution. Therefore, to make robust comparisons across platforms, we used order statistics \cite{David2003Order}. The RMA output was converted to a rank order. Next, we want to convert this rank order to the z-score of a log-normal distribution. We converting the rank to a percentile (for $N$ genes, divide by $N+1$), and then this percentile into a normal z-score. For later mathematical convenience, we used a biased estimator (normalize by $N$ not $N-1$) since then the Euclidean norm of each microarray gene expression is $N$. 

At this point, the natural dataset consisted of 601 microarrays with 20719 genes. Since we were interested in cellular identity, only transcription factors, transcription factor co-factors, or chromatin remodeling genes were kept (for short hand, referred to as transcription factors (TF) throughout the text) \cite{Zhang2012AnimalTFDB}, leaving 1715 TFs.

As explained in the main text, since continuous (sigmoidal input) attractor neural networks and discrete attractor neural networks are known to have the same stable fixed points \cite{Hopfield1984Neurons}, we used the binarized gene expression. We binarized the gene expression by setting a positive z-score to $+1$ and a negative z-score to $-1$. While this was mainly done for mathematical convenience, this is potentially biologically justified. Histone modifications (HM) either leave chromatin in an open, accessible configuration or a closed, inaccessible state \cite{Jenuwein2001Translating}. We found global HM data for embryonic stem cells (ESC), mouse embryonic fibroblasts (MEF), and neural progenitor cells (NPC) \cite{Mikkelsen2007Genome-wide,Meissner2008Genome-scale}. Consequently, we used the global HM data for these three cell fates and compared them to microarray TF expression levels. This allowed us to create a conditional probability distribution of each HM  for a given TF expression level (Figure 2B). We found a sharp cutoff (that coincides with a z-score of $0$) which distinguished TFs with the activating modification of histone 3 tri-methylation at lysine 4 (K4) from TFs with the inactivating modification of histone 3 tri-methylation at lysine 27 (K27), poised/bivalent TFs (both K4 and K27), and no HM (most likely DNA methylation). This shows that our mathematical assumption is justified by the HM data.

After the binarization of TF expression, all TFs that were not differentially expressed across cell fates (i.e. TFs that are always on / always off in every cell fate) were dropped, leaving 1337 TFs. The binarized TF expression for the 63 cell fates was found by first binarizing all 601 microarrays and then taking the majority vote for each cell state (with ties broken by averaging the continuous data). The final result was the binary expression state for 63 cell fates.

Microarrays for partially reprogrammed cells were on the Affymetrix GeneChip Mouse Genome 430 2.0 Array. The same procedure was used to convert raw microarray data to z-score expression. However, since different microarrays do not have the same genome coverage, the analysis comparing partially reprogrammed cells and natural cell fates used the $N=1329$ TFs common to both platforms.

Several self-consistency checks were performed on the data. First, the correlation matrix $A^{\mu\nu}$ (explained in main text and below)  was calculated for the original continuous data and for the binarized data (Figure S1). Both correlation matrices are consistent with each other showing binarization does not change the global correlations. Note that in the correlation matrix, cell fates have been grouped by tissue type, leading to a block diagonal form. Second, the expression state of all cell fates was constructed from multiple microarray experiments. These different experiments were compared with each other and were within 2 standard deviations (std equal to $1/\sqrt{N}\approx0.027$) for all cell fates. This  demonstrates that microarrays from multiple laboratories can be directly compared.

\subsection*{Landscape Model}

Here we give an overview of our epigenetic landscape model. The model is summarized in Table 1, and Text S1 provides a supplementary overview of attractor neural networks. 

\subsubsection*{State Space}

Each TF (labeled by $i$, $j$) can be in a state $S_i=\pm 1$ where $+1$ indicates the TF is active while $-1$ indicates it is inactive. A general cell state is given by $\textbf{S}$, an $N=1337$ dimensional vector. There are $p=63$ cell fates (labeled by $\mu$, $\nu$). In cell type $\mu$, the state of TF $i$ is given by $\xi_i^\mu$. The complete cell type data $\xi$ is a $p$ by $N$ matrix determined using our microarray data described above and these $\xi$ are the only biological input into the landscape. 

\subsubsection*{Full Landscape}

The complete landscape $H$ can be written as the following terms:
\begin{eqnarray}
H &=&  H_{basin} + H_{bias} + H_{culture} + H_{switch} \label{landscape}
\end{eqnarray}

Our landscape  assigns an ``energy'' to every global expression state. We emphasize that this energy does not correspond to physical energy consumption of ATP; instead it is an abstract energy that corresponds to stability and  developmental potential of cell fates. Each of the four terms has a simple interpretation (see Figure 1). The first term, $H_{basin}$, ensures that observed cell fates are valleys in our landscape (Figure 1A). The second term, $H_{bias}$,  describes biasing of specific TFs by experimentalists (not shown in Figure 1). The third term, $H_{culture}$, increases the radius and depth of  cell fates that are favored by the environment or culturing conditions (Figure 1B). Finally, in the presence of an external signal that gives rise to differentiation (ex. growth factors associated with differentiation), the fourth term, $H_{switch}$, opens a low energy path between the initial and final cell fates (Figure 1C).
\subsubsection*{Landscape Details: $H_{basin}$}

The gene expression profiles of naturally occurring cell fates must be minima of our landscape. This is ensured by the  landscape term 
\begin{eqnarray}
H_{basin} &=& -\frac{1}{2}\displaystyle\sum_{i=1}^N \displaystyle\sum_{j \neq i}^N S_i J_{ij} S_j 
\end{eqnarray}
In order to guarantee that cell fates are basins of attraction, we need to choose the ``effective interaction'' matrix, $J_{ij}$, which encodes how the $j$th TF influences the $i$th TF. Since we have highly correlated cell fates, we use the projection-method\cite{Kanter1987Associative}  (see Text S1 section ``Discrete, Projection Method''  for extended discussion on this choice), which defines the interaction matrix as:
\begin{equation}
J_{ij}=\frac{1}{N}\sum_{\mu=1}^{p} \sum_{\nu=1}^{p} \xi_i^\mu (A^{-1})^{\mu\nu} \xi_j^\nu
\end{equation}
where $\xi_i^\mu$ are the natural cell fates and $A^{-1}$ is the inverse of the correlation matrix between cell fates. Since our construction is based on correlations between gene expression profiles, $J_{ij}$ includes the effect of  ``indirect'' interactions  between TFs $i$ and $j$ that are mediated through other TFs (see Text S1 for additional mathematical explanation of this construction). While the current definition implies $J_{ij}$ is symmetric, this can easily be generalized to an asymmetric $J_{ij}$ (see later section Landscape vs Pseudo-Landscape for details).

\subsubsection*{Landscape Details: $H_{bias}$}

The term $H_{basin}$ ensures that all cell fates are global minima of the landscape. However, additional terms in the landscape are needed in order to incorporate key experimental features.

First, biologists can directly manipulate gene expression. For example, during the Yamanaka experiment,  the TFs $\emph{Pou5f1 (Oct4)}$, $\emph{Sox2}$, $\emph{Klf4}$, and $\emph{Myc}$ are overexpressed in fibroblasts. Mathematically, we represent the overexpression of TF $i$ by a local biasing field $B_i$ that ensures that $S_i=1$. Therefore the Yamanaka reprogramming protocol enters the landscape as:
\begin{equation}
H_{bias} =  -\displaystyle\sum_{i=1}^N  B_i S_i
\end{equation}
where for the Yamanaka protocol, $B_{\emph{Pou5f1}} = B_{\emph{Sox2}} =  B_{\emph{Klf4}} = B_{\emph{Myc}} \rightarrow \infty $ and for any other TF $i$, the field $B_i=0$.

\subsubsection*{Landscape Details: $H_{culture}$}

Currently, the basins of attraction $H_{basin}$ are all set to the same minima value. However, environmental signals (such as cell culture conditions) can stabilize and destabilize specific cell fates (see Figure 1B). This can be incorporated into our landscape by terms such as:
\begin{eqnarray}
H_{culture} &=&  - N\displaystyle\sum_{\mu=1}^p  b^\mu a^\mu \\
&=& - \displaystyle\sum_{i=1}^N C_i S_i \label{cult_full} 
\end{eqnarray}
where $b^\mu$ represents the culture biasing, and $a^\mu$ is the projection onto cell fate $\mu$. This bias can be equivalently expressed at the level of TFs by defining a culture bias, $C_i$, for the $i$th TF given by:
\begin{equation}
C_i = \displaystyle\sum_{\mu=1}^p \displaystyle\sum_{\nu=1}^p   b^\mu \left(A^{-1} \right)^{\mu\nu} \xi_i^\mu
\end{equation}
For example during the Yamanaka protocol, cells are cultured in conditions favorable to ESC, which is mathematically represented by $b^{ESC} >0$, while for all other cell fates $\mu$, $b^\mu=0$.

\subsubsection*{Landscape Details: $H_{switch}$}

During standard development, cells switch fates deterministically in response to external signals. We mathematically represent this using the term:
\begin{eqnarray}
H_{switch} &=&  -\frac{N}{2}\displaystyle\sum_{\mu=1}^p\displaystyle\sum_{\nu=1}^p m^\mu G^{\mu \nu} a^\nu  \label{H_switch_order} \\
&=&  -\frac{1}{2}\displaystyle\sum_{i=1}^N \displaystyle\sum_{j \neq i}^N S_i K_{ij} S_j \label{H_switch_K}
\end{eqnarray}
where  $m^\mu$ is the overlap on cell fate $\mu$, $a^\nu$ is the projection onto cell fate $\nu$, and the matrix $G^{\mu \nu}$ is the developmental signal matrix that is a dynamic entity and a function of developmental time and external signals. We can equivalently write this in terms of transcription factors using the gene-interaction matrix, $K_{ij}$, defined as:
\begin{equation}
K_{ij} =  -\frac{1}{N}\displaystyle\sum_{\mu=1}^p\displaystyle\sum_{\nu=1}^p \displaystyle\sum_{\rho=1}^p \xi^\mu_i G^{\mu \nu} \left(A^{-1} \right)^{\nu\rho} \xi_j^\rho  
\end{equation} 
where $\xi$ is the natural cell fate states, $G^{\mu\nu}$ is the developmental signal matrix, and $A^{-1}$ is the inverse correlation matrix. Since $G^{\mu\nu}$ is asymmetric, $K_{ij}$ is also asymmetric and explicitly breaks detailed balance (see later section Landscape vs Pseudo-Landscape for details).

We now explain the development signal matrix in more details. If $G^{\mu \nu} > 0$, this opens up a low energy path between cell fate $\nu$ and cell fate $\mu$. For example, during blood development, the common myeloid progenitor (CMP) can differentiate into either granulo-monocytic progenitors (GMP) or megakaryocyte-erythroid progenitors (MEP). The complicated external signals that induce switching from a CMP to GMP leads to $G^{GMP,CMP} >0 $ and all other $G^{\mu\nu}=0$, while the signals that induce switching from a CMP to MEP leads to $G^{MEP,CMP} > 0 $ and all other $G^{\mu\nu}=0$. We emphasize that this term is purely phenomenological and further research will be needed to directly connect the developmental biology signals (such as $TGF\beta$, $WNT$, etc) to the matrix elements  $G^{\mu \nu}$.

\subsubsection*{Dynamics}

We have uniquely defined the landscape $H$. However, there are multiple ways to implement dynamics on this landscape. In this paper, we are primarily interested in the behavior of the stable fixed points and not dynamical trajectories. Therefore, we follow the standard convention in the attractor neural network literature and update the network by random, asynchronous updates (Glauber dynamics) \cite{Amit1992Modeling}. Therefore, at each update, a random TF, $i$, is selected and updated according to the probability 
\begin{equation}
P[S_i(t+1)]=\frac{e^{\beta h_i(t) S_i(t+1)}}{e^{\beta h_i(t)} + e^{-\beta h_i(t)}}
\end{equation}
where $S_i$ is the expression state of the $i$th TF, $\beta$ is an effective noise parameter, $h_i$ is the local field, and $t$ is the time index. The local field $h_i$ is the gradient of the landscape (covariant derivative) defined for the full landscape $H$ as:
\begin{eqnarray}
h_i&=& \displaystyle\sum_{j\neq i}^N J_{ij} S_j + B_i  + C_i + \displaystyle\sum_{j\neq i}^N K_{ij} S_j 
\end{eqnarray}
where $J_{ij}$ is the basin-inducing interaction matrix, $B_i$ is the experimentally induced bias on the $i$th TF, $C_i$ is the culturing-condition specific bias on the $i$th TF, and $K_{ij}$ is the developmental interaction matrix.

We have introduced the effective noise parameter $\beta= 1/T$ (i.e. inverse temperature)  that controls the level of stochasticity resulting from biochemical noise. When $\beta \rightarrow \infty$, the update approaches a deterministic step function, while when $\beta \rightarrow 0$ each state is equally likely. Based on the currently available static genomic data, this update time cannot be directly related to biological time. Finally, we emphasize that since in this paper we are primarily concerned with the structure of the landscape, our results are independent of our choice of dynamics (see Text S1 for detailed discussion on dynamics).

\subsubsection*{Landscape vs Pseudo-Landscape}

Currently, the interactions between TFs, $J_{ij}$, are symmetric.  In real biology, this is unlikely to be true. We can introduce asymmetry into the interactions by randomly deleting interactions (for example Figure 2E Diluted). This asymmetry means that influence of TF $i$ on TF $j$ no longer equals the influence of TF $j$ on TF $i$. This asymmetry breaks detailed balance and implies a non-Lyapunov pseudo-potential \cite{Amit1992Modeling,Wang2010The-Potential, Wang2011Quantifying-Waddington} and has been shown to be an additional source of noise on the basins of attraction \cite{Amit1992Modeling}. 

We also note that the landscape term $H_{switch}$ is explicitly non-equilibrium and breaks detailed balance. Under one set of environmental conditions, $G^{\mu\nu}>0$ while $G^{\nu\mu}=0$ driving switching from $\nu \rightarrow \mu$, while under a different set of environmental conditions,  $G^{\nu\mu}>0$ while $G^{\mu\nu}=0$ driving switching from $\mu \rightarrow \nu$. Therefore, by including $H_{switch}$ we are actually making our landscape a pseudo-landscape (i.e. non-Lypanouv).

\subsection*{Simulations}

Here we include details of the simulations in this paper. For all simulations, we set $\beta=1/0.45 \approx 2.2$ and evolved the system for $100,000$ TF updates.

In Figure 2E, we demonstrate that we have basins of attraction. The initial conditions were created by taking the ESC expression vector and randomly flipping $15\%$ of the TFs.  After every $5000$ updates of asynchronous dynamics, burst errors were introduced by randomly flipping $2\%$ of TFs. For the asymmetric dilution, the standard interaction matrix $J_{ij}$ was created. Then $20\%$ of matrix entries were randomly set to $0$. 

In Figure 2F, we demonstrate that the landscape can deterministically switch between basins. The initial conditions were always the CMP expression vector. For signal 1, we set $G^{GMP,CMP}=0.5$ and all other $G^{\mu\nu}=0$. For signal 2, we set $G^{MEP,CMP}=0.5$ and all other $G^{\mu\nu}=0$.

\subsection*{Spurious Attractors}
Here we provide more details on spurious attractors and hybrid cell fates. As explained in more detail in Text S1, for the traditional Hopfield model, these spurious attractors take the form of odd-majority vote mixtures \cite{Amit1992Modeling} (i.e. majority vote at each TF of $3,5,7,\ldots$ of the $\xi_i^\mu$). The projection method also has the additional spurious attractors of any linear combination of $\xi_i^\mu$ that spans the discrete state space (see geometric interpretation given in Text S1)\cite{Kanter1987Associative}. For convenience, we use the word hybrid as the collective term for either majority vote mixtures or linear combinations of existing cell fates.

As discussed in the main text, the prediction of spurious attractors in the projection method inspired us to reexamine data on existing partially reprogrammed cells. Surprisingly, we found that partially reprogrammed cells could be thought of as hybrids of existing cell fates. However, we are currently only able to obtain qualitative agreement between partially reprogrammed cells and the predicted nature of the spurious attractors. While it is known that the projection method retains these odd-majority vote mixtures spurious attractors, the correlations between states implies these spurious attractors may no longer be symmetric mixtures. However, the exact nature of these spurious attractors is not known and will be explored in future work.

\section*{Acknowledgments}

We thank members of the Mehta Group, Collins lab, and Laertis Ikonomou, Katherine Benson, Darrell Kotton, and other members of Boston University Center for Regenerative Medicine (CReM) for stimulating discussions. In addition, we thank Laertis Ikonomou, Darrell Kotton, and Kristian Moss Bendtsen for a detailed reading of an earlier version of the manuscript.

%


\bibliographystyle{plos2009}

\clearpage{}

\begin{figure}[!ht]
\begin{center}
\includegraphics[width=5in]{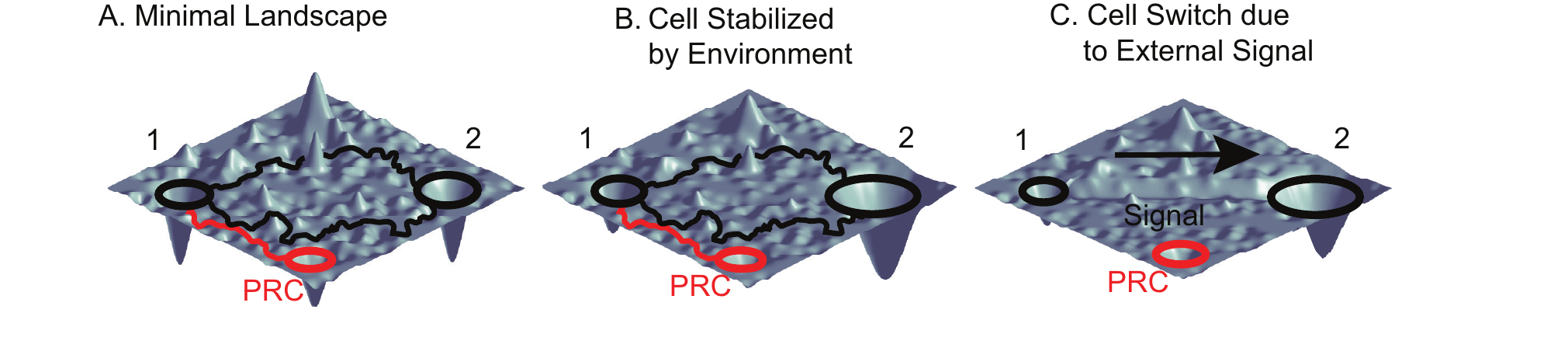}
\end{center}
\caption{
{\bf Phenotypic Landscape.}  These are illustrative cartoons of the cell fate attractor landscape. (A) The minimal cellular identity landscape. Each cell fate is a basin of attraction (black circles). Reprogramming between different cell fates (1 and 2) can occur probabilistically via different trajectories (black paths). Partially reprogrammed cells (PRC) exist as smaller, spurious, basins of attraction (red circle) that can be experimentally observed by reprogramming experiments (example trajectory in red). (B) Same cellular identity landscape in the presence of a stabilizing environment (ex. favorable culturing medium) for cell fate 2. The environment increases the radius and depth of the cell fate 2 basin of attraction. (C) Landscape in the presence of an external signal that gives rise to differentiation from cell fate 1 to cell fate 2 (ex. growth factors associated with differentiation). Notice the low energy path between the cell fates that drives switching from cell fate 1 to cell fate 2.}
\label{Figure_label_1}
\end{figure}

\begin{figure}[!ht]
\begin{center}
\includegraphics[width=5in]{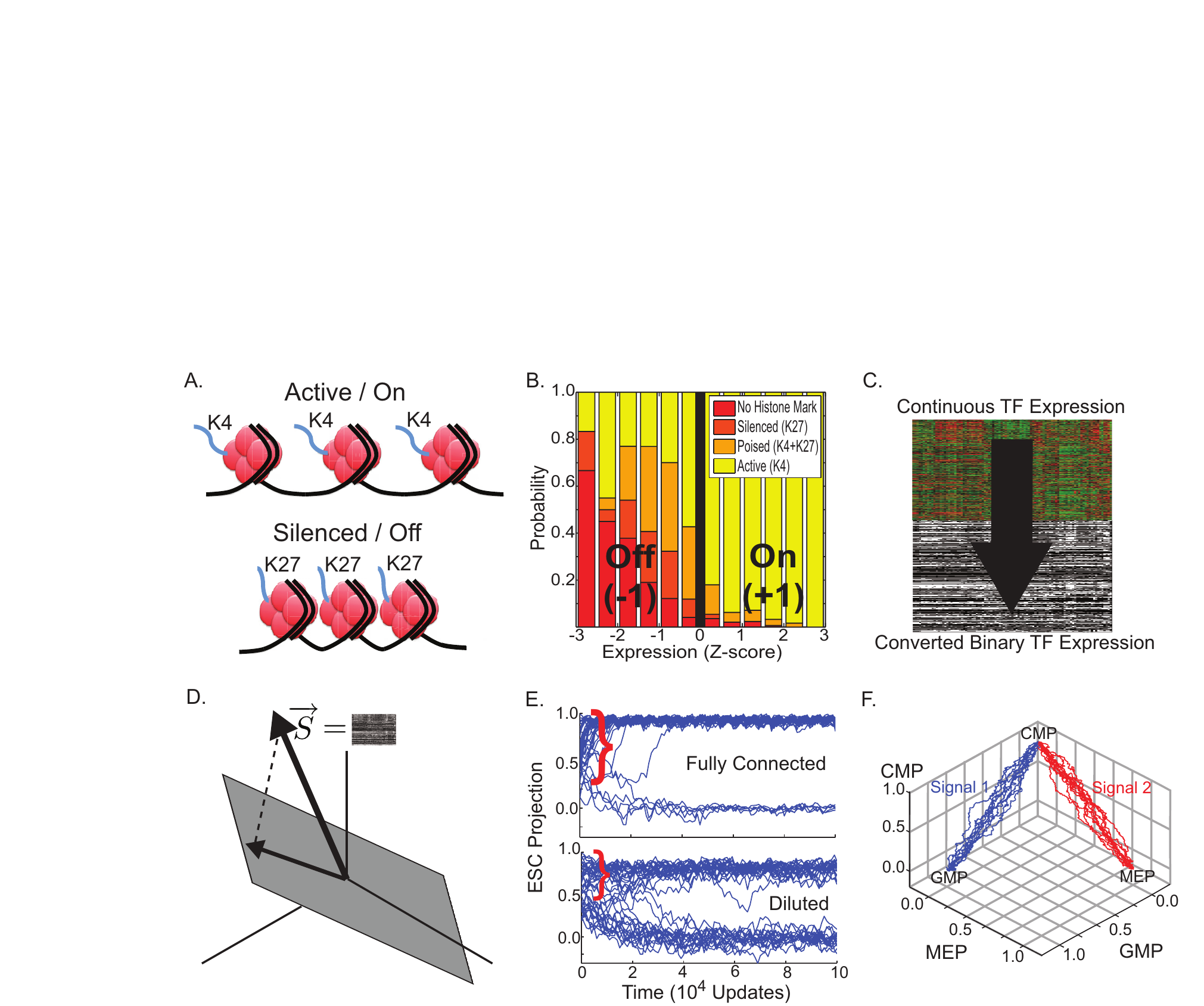}
\end{center}
\caption{
{\bf Overview of model.}  (A) Histone 3 tri-methylation at lysine 4 (K4)  is associated with active genes,  while histone 3 tri-methylation at lysine 27 (K27) is  associated with repressed genes. (B) Conditional probability distribution of histone modification (HM) given transcription factor (TF) expression levels derived by comparing microarray data with HM data from \cite{Mikkelsen2007Genome-wide,Meissner2008Genome-scale}. Notice the sharp threshold (black line) between expression levels of active and inactive TFs. (C) For mathematical convenience, we take the continuous TF expression levels and convert it to binary states (z-score $>=0$ to $+1$ and z-score $<0$ to $-1$). This binarization is consistent with the result from (B). (D) An arbitrary state is represented by a vector $\vec{S}$ of $\pm 1$, with each dimension in the vector space  representing the state of a TF. The natural cell fates form a subspace (gray plane). The landscape model is based on the orthogonal projection of the TF state onto this subspace. (E) The dynamics of the landscape model for different initial conditions for a fully connected interaction matrix $J_{ij}$ and a diluted (non-equilibrium) interaction matrix where 20\% of interactions have been randomly deleted. Plot shows the projection of $S$ on embryonic stem cells (ESC) as function of time. Notice the large basins of attraction  (red bracket). Parameters used were $\beta=2.2$ and burst errors of $2\%$ every $5000$ spin updates.  (F) Simulations showing how a  common myeloid progenitor (CMP) can differentiate into either granulo-monocytic progenitors (GMP) or megakaryocyte-erythroid progenitors (MEP) in response to two distinct external signals. All trajectories used $\beta=2.2$. For signal 1, we set $G^{GMP,CMP}=0.5$ and all other $G^{\mu\nu}=0$. For signal 2, we set $G^{MEP,CMP}=0.5$ and all other $G^{\mu\nu}=0$.} 
\label{Figure_label_2} 
\end{figure}

\begin{figure}[!ht]
\begin{center}
\includegraphics[width=4in]{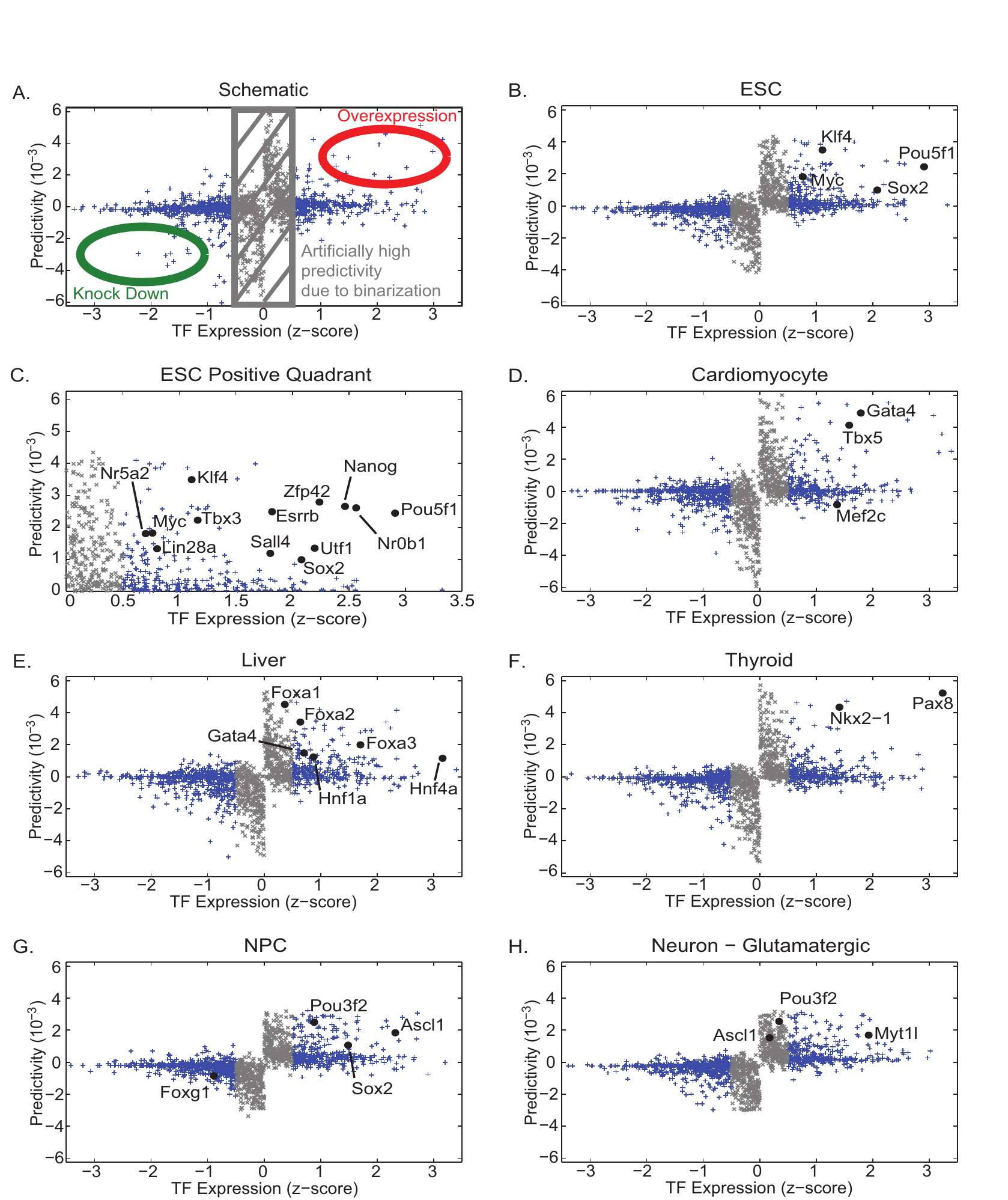}
\end{center}
\caption{
{\bf Identifying reprogramming candidates.}  For a given cell fate, we plot every differentially expressed transcription factor's (TF) predictivity (aka energy projection-contribution, $\eta_i^\mu$) vs TF expression level (z-score normalized). Unless otherwise stated all existing reprogramming protocols to a given cell fate are labeled. (A) Schematic illustrating predictivity vs expression level plots. The large positive (negative) predictivity and large positive (negative) gene expression TFs are candidates for over expression (knock out) in a reprogramming protocol. The TFs with z-score between $-0.5$ and $0.5$ are highlighted in gray because Figure 2B suggests these TFs predictivity may be prone to extra noise induced by the data discretization. (B) Embryonic stem cell, ESC (induced pluripotent stem cells, iPSC). Original Takahashi and Yamanaka factors \emph{Pou5f1} (\emph{Oct 4}), \emph{Sox2}, \emph{Klf4}, and \emph{Myc} \cite{Takahashi2006Induction}. (C) Inset of ESC positive predictivity and gene expression. \emph{Zfp42} (\emph{Rex1})\cite{Masui2008Rex1/Zfp42} and \emph{Nr0b1} (\emph{Dax1})\cite{Khalfallah2009Dax-1} are pluripotency markers that are not necessary to overexpress for reprogramming, while combinations of the remaining labeled TFs have been successfully used in reprogramming protocols \cite{Gonzalez2011Methods}. (D) Heart (induced cardiomyocytes, iCM) \cite{Ieda2010Direct}. (E) Liver (induced hepatocytes, iHep). There are two published protocols. One protocol used \emph{Hnf4a} plus any of \emph{Foxa1}, \emph{Foxa2}, or \emph{Foxa3} \cite{Sekiya2011Direct} while another used \emph{Gata4}, \emph{Foxa3}, \emph{Hnf1a}, and deletion of \emph{p19Arf} \cite{Huang2011Induction}. \emph{p19Arf} was not differentially expressed  in our microarrays and is not shown. (F) Thyroid \cite{Antonica2012Generation}. (G) Neural Progenitor Cells, NPC (induced NPC, iNPC) used \emph{Pou3f2} (\emph{Brn2}), \emph{Sox2}, and \emph{Foxg1} \cite{Lujan2012Direct}. With our microarrays we find that \emph{Foxg1} is not predictive for NPC but is predictive of neural stem cells (NSC) (see Figure S3). (H) Neurons (induced neuron, iN) \cite{Vierbuchen2010Direct}. The reprogramming protocol used a combination of factors that were known to be important to ether mature neurons (\emph{Myt1l}) or NPCs (\emph{Pou3f2}, \emph{Ascl1}). (G) shows that \emph{Pou3f2} and \emph{Ascl1} are predictive of NPCs.
}
\label{Figure_label_3}
\end{figure}


\clearpage{}


\begin{table}[!ht]
 \caption{\textbf{Mathematical model of cell identity landscape} }
 \begin{tabular}{| p{5cm} | p{4cm}  | p{5cm} | }
 \hline
Landscape Term: Index Notation & Landscape Term: Matrix Notation (dim.) & Biological Interpretation  \\
\hline
$H = H_{basin} + H_{bias} + H_{culture} + H_{switch}  $  & $ $ &  Total landscape. \\
\hline
$H_{basin} = -\frac{1}{2}\displaystyle\sum_{i=1}^N \displaystyle\sum_{j \neq i}^N S_i J_{ij} S_j $ &  $H_{basin} = -\frac{1}{2}  \textbf{S}^\textsf{T} J  \textbf{S}$& Produces cell basins of attraction.  \\
\hline
$H_{bias} =  -\displaystyle\sum_{i=1}^N  B_i S_i$ & $H_{bias} =  - \textbf{B}^\textsf{T} \textbf{S}$ & External control of individual genes, i.e. inducible expression.  \\
\hline
$H_{culture} = -N\displaystyle\sum_{\mu=1}^p  b^\mu a^\mu$ &  $H_{culture} = -N \textbf{b}^\textsf{T} \textbf{a}$ & External control of specific cell basins, i.e. culturing conditions.\\
\hline
$H_{switch} = -\frac{N}{2}\displaystyle\sum_{\mu=1}^p\displaystyle\sum_{\nu=1}^p m^\mu G^{\mu \nu} a^\nu$ & $H_{switch} = -\frac{N}{2} \textbf{m}^\textsf{T} G  \textbf{a}$ & Cell switching by signals, i.e. \emph{in vivo} development.  \\
\hline
$N$ &   & Number of TFs, labeled by $i$, $j$. In this paper $N=1337$.\\
\hline
$p$ &  & Number of cell fates, labeled  by $\mu$, $\nu$. In this paper $p=63$.\\
\hline
$S_i$ & $\textbf{S}$ \hspace{1 mm} ($N$ x $1$) & State ($\pm1$) of $i^{th}$ TF.  \\
\hline
$\xi_i^\mu$ & $\xi$ \hspace{1 mm} ($p$ x $N$) & State ($\pm1$) of $i^{th}$ TF in cell fate $\mu$.  \\
\hline
$A^{\mu\nu} = \frac{1}{N}\displaystyle\sum_{i=1}^N \xi_i^\mu \xi_i^\nu$ & $A= \frac{1}{N} \xi \xi^\textsf{T}$ \hspace{1 mm} ($p$ x $p$) & Correlation between cell fate $\mu$ and $\nu$.  \\
\hline
$J_{ij} = \frac{1}{N}\displaystyle\sum_{\mu=1}^{p} \displaystyle\sum_{\nu=1}^{p} \xi_i^\mu (A^{-1})^{\mu\nu} \xi_j^\nu$ &$J = \frac{1}{N} \xi^\textsf{T} A^{-1}\xi $  \hspace{1 mm} ($N$ x $N$)& Interaction strength between $i$ and $j$. \\
\hline
$B_{i}$ & $\textbf{B}$ \hspace{1 mm} ($N$ x $1$)  & External control of $i^{th}$ TF.  \\
\hline
$b^{\mu}$ &  $\textbf{b}$ \hspace{1 mm} ($p$ x $1$) & External control of $\mu^{th}$ cell fate.  \\
\hline
$m^{\mu} = \frac{1}{N}\displaystyle\sum_{i=1}^N \xi_i^\mu S_i$ & $\textbf{m} = \frac{1}{N} \xi \textbf{S}$ \hspace{1 mm} ($p$ x $1$) &  Overlap of $\textbf{S}$ on cell fate $\mu$.  \\
\hline
$a^{\mu} = \displaystyle\sum_{\nu=1}^p (A^{-1})^{\mu\nu}m^\nu = \displaystyle\sum_{i=1}^N \eta_i^\mu S_i$ &  $\textbf{a} = A^{-1} \textbf{m} =  \eta \textbf{S}$ \hspace{1 mm} ($p$ x $1$) &Projection of $\textbf{S}$ on cell fate $\mu$.  \\
\hline
$\eta^{\mu}_i=  \frac{1}{N}\displaystyle\sum_{\nu=1}^p (A^{-1})^{\mu\nu}\xi_i^\nu$ & $\eta =  \frac{1}{N} A^{-1}\xi$  \hspace{1 mm} ($p$ x $N$) & Predictivity of $i^{th}$ TF in cell fate $\mu$. \\
\hline
$G^{\mu\nu}$ & $G$  \hspace{1 mm} ($p$ x $p$)  & Signal dependent coupling that drives cell fate $\nu$ to cell fate $\mu$\\
\hline
\end{tabular}
 \begin{flushleft}
This table provides a summary of the landscape model and the biological interpretation of each term. The first column is written in index notation, while the second column is the same term in matrix notation with the dimension of the term given in parenthesis. If no dimension is listed, the term is a single number.
\end{flushleft}
 \end{table}

\newpage{}

 \begin{table}[!ht]
 \caption{\textbf{Partially reprogrammed cells as spurious attractors.} }

 \begin{tabular}{| p{2cm} | p{1cm} | p{1cm} | p{7cm} | }
 \hline
Cell line & Start & Goal & Highest projecting states (projection)  \\
 \hline

1A2 \cite{Sridharan2009Role} &MEF & ESC & 	\textbf{ESC (0.178)}, MSC (0.158), myoblast (0.142), MEP (0.129), blood vessel (0.113), keratinocyte (0.112), medullary thymic epithelial (-0.111), adipose - brown (-0.117), NK (-0.130), CMP (-0.138)    \\
\hline
1B3 \cite{Sridharan2009Role} & MEF & ESC & \textbf{ESC (0.222)}, \textbf{MSC (0.161)}, blood vessel (0.139), myoblast (0.138), GMP (0.127), kidney (0.111), MEP (0.107), cornea (0.107), NK (-0.129)   \\
\hline
BIV1+ \cite{Mikkelsen2008Dissecting} & B Cell & ESC & \textbf{myoblast (0.181)}, \textbf{prostate (0.164)}, MSC (0.154), MEP (0.138), keratinocyte (0.136), cornea (0.125), ESC (0.111), intestine - Paneth cell (-0.111), CMP (-0.122)    \\
\hline
BIV1- \cite{Mikkelsen2008Dissecting} & B Cell & ESC & \textbf{ESC (0.382)}, \textbf{EpiSC (0.184)}, \textbf{MEP (0.160)}, myoblast (0.145), NSC (-0.108), T Cell (-0.115), skeletal muscle (-0.117), CMP (-0.154)   \\
\hline
MCV6 \cite{Mikkelsen2008Dissecting} & MEF & ESC & MEP (0.155), myoblast (0.150), ESC (0.149), keratinocyte (0.145), CLP (0.107), GMP (0.107), cornea (0.107), CMP (-0.130)  \\
\hline
MCV8 \cite{Mikkelsen2008Dissecting} & MEF & ESC & \textbf{ESC (0.203)}, \textbf{MEP (0.191)}, \textbf{myoblast (0.160)}, cornea (0.119), prostate (0.113), skeletal muscle (-0.141), CMP (-0.142)   \\
 \hline
 \end{tabular}
 \begin{flushleft}
 Partially reprogrammed cell lines (first column) and their significant projections (2 std above noise or $|a|>0.106$)  onto ``natural'' cell fates based on microarray data.  Bold indicates 3 std above noise or $|a|>0.159$. Abbreviations: CLP, Common Lymphoid Progenitor; CMP, Common Myeloid Progenitor; EpiSC, epiblast stem cell; ESC, embryonic stem cell; GMP, Granulocyte-Monocyte Progenitor; MEF, mouse embryonic fibroblast; MEP, Megakaryocyte-Erythroid Progenitor; MSC, Mesenchymal stem cells; NK, Natural Killer cells; NSC, neural stem cells.
 \end{flushleft}
 \end{table}

\newpage{}

\section*{Supporting Information Legends}

\paragraph{Text S1}
\textbf{Attractor Neural Networks: Additional Details.}
This supplementary text provides extended background details on Hopfield attractor neural networks but presents no new research findings. The sections are: (A) Discrete, Standard Hopfield. (B) Continuous, Standard Hopfield. (C) Continuous Gene Expression. (D) Discrete as Limit of Continuous. (E) Discrete, Projection Method.

\vspace{3mm}

\paragraph{Figure S1}
\textbf{Cell fate correlation matrices.}
(A) Correlation matrix between cell fates for continuous data. (B) Correlation matrix for binarized data.

\vspace{3mm}

\paragraph{Figure S2}
\textbf{Projection of a random vector on a given cell fate.}  Ten thousand binarized random vectors were created in MATLAB and projected onto the Õcellular sub-spaceÕ. The histogram shows the distribution of the projections. The red line is a Gaussian fit to the histogram. The mean is practically zero while the standard deviation is 0.053.

\vspace{3mm}

\paragraph{Figure S3}
\textbf{Predictivity vs Expression for NSC.}  Same type of plot as Figure 3. Labeled TFs are part of reprogramming protocol to NPC \cite{Lujan2012Direct}. This illustrates that \emph{Foxg1} is predictive for NSC, even though it is not for NPC.

\vspace{3mm}

\paragraph{Table S1}
\textbf{Classifying Top ESC Reprogramming Candidates.}  Table has top 50 embryonic stem cell (ESC) reprogramming candidates (as ranked by z-score times predictivity, $\eta_i^\mu$). Classification of each TF is either justified by paper citation or GO Process term.

\vspace{3mm}

\paragraph{Table S2}
\textbf{Examining Yamanaka Factors in Detail.}  Here we reexamine the Yamanaka transcription factors (TFs) in light of our model.

\vspace{3mm}

\paragraph{File S1}
\textbf{Microarray Sources}.  List of all microarrays used in this paper.

\vspace{3mm}

\paragraph{File S2}
\textbf{TF Z-Score}. The z-score gene expression for each TF of natural cell fates in this paper. This data is post RMA normalization and averaging over multiple replicates for each natural cell fate.

\vspace{3mm}

\paragraph{File S3}
\textbf{TF Predictivity}.  The predictivity for each TF and cell fate in this paper.

\vspace{3mm}

\paragraph{File S4}
\textbf{Partially Reprogrammed Cells Z-Score}.    Partially Reprogrammed Cells Z-Score.  The z-score gene expression for each TF of partially reprogrammed cell fate. This data is post RMA normalization and averaging over multiple replicates for each partially reprogrammed cell fate.
\vspace{3mm}

\paragraph{File S5}
\textbf{Overexpression Candidates}.   Top overexpression candidates to reprogram to various cell fates.

\vspace{3mm}

\paragraph{File S6}
\textbf{Knock-Out Candidates}.   Top knock-out candidates to reprogram to various cell fates.

\end{document}


\title{Supplementary Information: Epigenetic landscapes explain partially reprogrammed cells and identify key reprogramming gene}

\author{Alex~H.~Lang}
\affiliation{Physics Department, Boston University, Boston, MA, USA}
\affiliation{Center for Regenerative Medicine, Boston University, Boston, MA, USA}

\author{Hu~Li}
\affiliation{Department of Biomedical Engineering, Boston University, Boston, MA, USA}
\affiliation{Wyss Institute for Biologically Inspired Engineering, Harvard University, Boston, MA, USA}

\author{James~J.~Collins}
\affiliation{Department of Biomedical Engineering, Boston University, Boston, MA, USA}
\affiliation{Wyss Institute for Biologically Inspired Engineering, Harvard University, Boston, MA, USA}
\affiliation{Howard Hughes Medical Institute, Boston, MA, USA}
\affiliation{Center for BioDynamics, Boston University, Boston, MA, USA}

\author{Pankaj~Mehta}
\affiliation{Physics Department, Boston University, Boston,  MA, USA}
\affiliation{Center for Regenerative Medicine, Boston University, Boston, MA, USA}

\maketitle

\tableofcontents

\section{Attractor Neural Networks: Additional Details}
This supplementary text gives a  brief introduction to Hopfield neural networks \cite{Hopfield1982Neural,Amit1985Spin-glass} and how they can be adapted to study epigenetic landscapes. We begin by  reviewing  the basic principles underlying the original Hopfield neural network. We then show how to generalize this to continuous  spins\cite{Hopfield1984Neurons} as well as discrete spins  with correlated cell fates \cite{Kanter1987Associative} (projection method). For an in-depth introduction to neural networks, please see the beautiful book by  Amit \cite{Amit1992Modeling}.

\subsection{Discrete, Standard Hopfield}
There are $N$ genes and each gene $i$ is either on or off, with the output denoted by $S_i=\pm1$. Alternatively, we could use the variables $\widetilde{S} = \frac{1}{2}(S+1)=1,0$ with the corresponding substitutions in all equations below.

The input to a given gene $i$ is denoted by the local field
\be
h_i = \displaystyle\sum_{j\neq i}^N J_{ij}S_j+ B_i
\ee
where $J_{ij}$ is the interaction between gene $i$ and gene $j$ and $B_i$ is the external (i.e interaction independent) bias of gene $i$. Both $J_{ij}$ and $B_i$ are assumed to be independent of $S_i$.

The landscape $H$ is given by
\beq
H &=& -\frac{1}{2}\displaystyle\sum_{i=1}^N \displaystyle\sum_{j\neq i}^N S_i J_{ij}S_j -\displaystyle\sum_{i=1}^N B_i S_i \\
 &=& -\frac{N}{2} \displaystyle\sum_{\mu=1}^p \left(m^\mu\right)^2 - N\displaystyle\sum_{\mu=1}^p b^\mu m^\mu \label{hop_mag}
\eeq
where in equation \ref{hop_mag} we have introduce the order parameter for the overlap (dot product or ``magnetization'') of a spin configuration with a given cell fate $\mu$ as $m^\mu$ and also introduced the cell fate bias $b^\mu$. The overlap is defined in terms of the cell fate vectors $\xi_i^\mu$ as:
\be
m^\mu = \frac{1}{N} \displaystyle\sum_{i=1}^N \xi^\mu_i S_i
\ee

To prove that $H$ is a Lypanov function (i.e. has stable equilibrium states and follows the standard definition of an ``energy''), it is necessary to show that $H$ is a decreasing function and bounded below.  To do so, consider flipping a single $S_i$. The resulting change in $H$ is
\be
\Delta H = -\frac{1}{2}\left[ \displaystyle\sum_{j\neq i}^N J_{ij}S_j + \displaystyle\sum_{j\neq i}^N S_j J_{ji} + B_i \right]  \Delta S_i
\ee

When we have symmetric interactions, $J_{ij}=J_{ji}$, this simplifies to 

\be
\Delta H = -\left[ \displaystyle\sum_{j\neq i}^N J_{ij}S_j + B_i \right]  \Delta S_i = - h_i \Delta S_i
\ee

To determine the sign of $\Delta H$ we need the relation between $h_i$ and $\Delta S_i$. For deterministic (stochastic) dynamics, as long as $\Delta S_i$ and $h_i$ are always (usually) the same sign, we always (usually) have $\Delta H < 0$. Therefore, any set of dynamics that stochastically matches the sign of $\Delta S_i$ and $h_i$ will lead to $H$ being a Lypanov function. This implies that any choice of dynamics leads to the same stable fixed points, but may give rise to different trajectories, limit cycles, and sizes of basins of attraction for fixed points, see Amit\cite{Amit1992Modeling} section 2.2 and 3.5 for a detailed analysis. Therefore, in this paper we focus on predictions that are independent of the exact dynamics. This is equivalent to thinking about the stationary properties of the model.

We will follow the standard convention for neural networks and physics and implement Glauber dynamics which is an asynchronous, stochastic update rule. In this update scheme, at each time step, one gene is selected at random and probabilistically updated according to its local field

\begin{equation}
P[S_i(t+1)]=\frac{e^{\beta h_i(t) S_i(t+1)}}{e^{\beta h_i(t)} + e^{-\beta h_i(t)}}
\end{equation}

with $h_i$ defined above (or equivalently $h_i=-\frac{\partial H}{\partial S_i}$) and $t$ time measured in discrete updates. Also, $\beta = 1/T$ is the inverse temperature and characterizes the slope of the sigmoid function. When $\beta \rightarrow \infty$, the sigmoid approaches a deterministic step function, while when $\beta \rightarrow 0$ each state is equally likely. 

Now we need to specify the gene interaction $J_{ij}$ and establish the global minima of the system. There are $p$ cell fates and the state of gene $i$ in cell fate $\mu$ is given by $\xi_i^\mu$. The gene interaction is a correlation based interaction and in the standard Hopfield neural network it is defined as

\be
J_{ij}=\frac{1}{N}\displaystyle\sum_{\mu=1}^p \xi_i^\mu \xi_j^\mu
\ee

In the standard Hopfield network, the cell fates have two assumptions. First, each cell fate is assumed to on average be unbiased (i.e. equal number of positive and negative spins)

\be
\frac{1}{N}  \displaystyle\sum_{i=1}^N \xi^\mu_i \approx 0
\ee

and second every pair of cell fates is approximately orthogonal

\be
\frac{1}{N} \displaystyle\sum_{i=1}^N \xi^\mu_i \xi^\nu_i \approx \mathcal{O}\left( \frac{1}{\sqrt{N}} \right)
\ee

These two assumptions can be relaxed in extensions of the standard Hopfield neural network, see later sections for one example (the projection method) that can incorporate correlated cell fates.

Now we can prove that each cell fate is a global minima of the landscape. For no external fields, the landscape can be written as:

\be
H = -\frac{1}{2}\displaystyle\sum_{i=1}^N \displaystyle\sum_{j\neq i}^N S_i J_{ij}S_j  = -\frac{N}{2} \displaystyle\sum_{\mu=1}^p \left( \frac{1}{N} \displaystyle\sum_{i=1}^N \xi^\mu_i  S_i \right)^2  + \frac{1}{2N} \displaystyle\sum_{i=1}^N \displaystyle\sum_{\mu=1}^p S_i \xi^\mu_i \xi^\mu_i S_i
\ee

This can be rewritten in terms of the overlap as:

\be
H = -\frac{N}{2} \textbf{m}^2 + \frac{1}{2}p
\ee

Then as long as $N$ is large compared to $p$, whenever we are in a given cell fate the energy is $H=-N/2$ and this is the lowest bound since $\textbf{m}^2 \le 1$. We have shown that for $p\ll N$, $H$ is a decreasing, bounded function and hence is a Lypanov function. When $p$ and $N$ are both large, a full replica calculation shows that $H$ remains a Lypanov function\cite{Amit1985Storing}.

While we have established that the landscape is a Lypanov function, we also need to examine the dynamical stability of the cell fates and the existence of spurious attractors. In the absence of stochastic update noise ($\beta \rightarrow \infty$), we can examine the signal-to-noise ratio of the cell fates. If a state is dynamically stable, one needs $S_i h_i >0$. When the state is in a given cell fate (without loss of generality assume cell fate $1$), we have that

\be
\xi^1_1 h_1 = \frac{1}{N} \displaystyle\sum_{j\neq i}^N \displaystyle\sum_\mu^p  \xi^1_1 \xi^\mu_1 \xi^\mu_j \xi^1_j
\ee

which can be broken into a signal term (first term) and noise term (second term) as follows:

\be
\xi^1_1 h_1 = \frac{N-1}{N} +  \frac{1}{N} \displaystyle\sum_{j\neq i}^N \displaystyle\sum_{\mu\neq 1}^N \xi^1_1 \xi^\mu_1 \xi^\mu_j \xi^1_j
\ee

For large $N$, the signal term approaches $1$. We can evaluate the noise term by recognizing that it is an unbiased sum of $(N-1)(p-1) \approx Np$ random steps, and therefore has mean $0$ and standard deviation $\sqrt{pN}$, giving us

\be
\xi^1_1 h_1 = 1 + \mathcal{O}\left( \sqrt{\frac{p}{N}} \right)
\ee

Therefore as long as $N$ is much larger than $p$, every cell fate is a fixed point. This rough signal-to-noise argument can be made more rigorous by a spin-glass replica calculation \cite{Amit1985Storing} which finds that cell fates are stable (in the case $\beta \rightarrow \infty$) as long as the ratio of $p/N$ is less than $0.138$. 

Here is an intuitive argument of why the landscape must be rugged, which implies the scaling of stable states with $N$. From looking at small systems, a naive guess would be that the number of stable states should scale with the size of the state space $2^N$. This scaling could be achieved if each minima occurred when a single TF state is turned on while all the other TFs are off. However, this implies that each minima is only marginally stable; any spin flip will move the state out of the minima. In order to have a basin of attraction, more TFs are needed to determine the minima. A simple error correction or redundancy could be implemented by using $r$ redundant TFs, but this would require exponentially more states $r^N$.  Instead, stable states could be determined by overlapping sets of TFs, as in the Hopfield neural network. This form of error-correction leads to frustration and Gaussian noise between the stable states, hence the scaling of stable states with $N$ and not $2^N$.

An unavoidable consequence of the non-linearity (ruggedness) of the Hopfield network is that in addition to the desired attractors (the input cell fates), there are additional spurious, metastable, attractors. There are a variety of spurious attractors, but the most common are symmetric mixtures of odd states \cite{Amit1985Spin-glass}, for example without loss of generality we can make a spurious state with the first three cell types, $S_{spur}=\text{major}\left(\xi_1+\xi_2+\xi_3\right)$,  where $\text{major}$ stands for majority vote (equivalently the sign function) at each spin. The most common spurious attractor are symmetric mixtures of 3 states (as in the example above). A signal-to-noise analysis can also be done to establish that these spurious attractors are stable attractors, but with a smaller basin of attraction than the input cell fates (see Amit 4.3 for details \cite{Amit1992Modeling}).

\subsection{Continuous, Standard Hopfield}
The previous section describes the basic ideas of Hopfield neural networks. Here, we show how discrete Hopfield neural networks can be considered a limiting case of continuous differential equations of gene expression.  We start by defining continuous spins,  $\S_i$, that can take on real number between $-1$ and $1$. For continuous dynamics, we must modify the dynamics of the corresponding local field. In particular, if the local field decays in time with a time constant $\tau_i$
we have
\be
\frac{dh_i}{dt} = \displaystyle\sum_{j \neq i}^N J_{ij} \S_j + B_i - \tau_i^{-1}h_i
\ee
where the $J_{ij}$ are the same as in the discrete case and the spin $\S_i$ is related to the local field by some monotonic function $\S_i = g_i\left[ h_i \right]$.

Now the landscape is given by
\be
H = -\frac{1}{2} \displaystyle\sum_{i=1}^N \displaystyle\sum_{j \neq i}^N \S_i J_{ij} \S_j -\displaystyle\sum_{i=1}^N B_i \S_i +\displaystyle\sum_{i=1}^N \tau_i^{-1} \int_{-1}^{\S_i} g_i^{-1}\left[ \S \right] d\S
\ee

where the first two terms are the same as in the discrete case while the third is the new term for continuous only. Taking derivatives with respect to time gives us

\be
\frac{dH}{dt} = -\displaystyle\sum_{i=1}^N \frac{d\S_i}{dt} \left( \displaystyle\sum_{j  \neq i}^N J_{ij} \S_j + B_i - \tau_i^{-1}h_i \right) = -\displaystyle\sum_{i=1}^N \frac{d\S_i}{dt} \frac{dh_i}{dt}
\ee

Then since $h_i = g_i^{-1}\left[ \S_i \right]$, we can relate the derivative of $h_i$ to the derivative $\S_i$. Then using the fact that $g_i$ is monotonically increasing we can show that the change in $H$ is always negative:
\be
\frac{dH}{dt} = -\displaystyle\sum_i^N g_i^{-1}\left[\S_i \right] \left( \frac{d\S_i}{dt} \right)^2 \le 0
\ee

The decrease in $H$ along with the fact that $H$ is bounded below, shows that we have a Lypanuv function. It is easy to see that every discrete stable point is also a stable point in the continuous model; however, the continuous Hopfield neural networks can have additional stable points.

\subsection{Continuous Gene Expression}

A popular approach to model gene interactions is based on the genetic toggle switch\cite{Gardner2000Construction} and represents gene interactions by a Hill function. For now, we will use the general variable $\widetilde{\sigma} \in [\sigma_{min},\sigma_{max}]$. 

In the most general case, we have that

\be
\widetilde{\sigma}_i = \text{sign}(h_i) \frac{a_i |h_i|^{n_i}}{k_i^{n_i}+|h_i|^{n_i}} + b_i
\ee

where the input $h_i$ is in the range $[-\infty, \infty]$ and the output $\sigma_i$ is in the range $[-a_i+b_i, a_i +b_i]$. 

If we rescale every gene by its dynamic range and center the Hill function at zero, we get that $\widetilde{\sigma} = \S \in [-1,1]$ and

\be
\S_i = \text{sign}(h_i) \frac{|h_i|^{n_i}}{k_i^{n_i}+|h_i|^{n_i}} 
\ee
Using the function above for $\S_i = g_i\left[ h_i \right]$ allows one to relate continuous Hopfield neural networks to gene expression using Hill coefficients.

\subsection{Discrete as Limit of Continuous}
How can we relate the continuous model of gene expression to the previous discrete model? There are two limits. First, if we take the discrete time limit with the update time much greater than the input memory, we get
\be
h_i(t+1) = \displaystyle\sum_{j  \neq i}^N J_{ij}S_j(t) + B_i
\ee

Second, in the genetic toggle switch language, when the cooperativity is large $n\gg 1$, then $S_i\rightarrow\pm1$. This gives us a deterministic, discrete model of gene expression. If we introduce stochasticity through Glauber dynamics, we completely recover the discrete Ising model of gene expression.

\subsection{Discrete, Projection Method}

The standard Hopfield attractor neural network assumes that the ``memories'' (cell fates) have nearly no correlations amongst themselves. However, cell fates are highly correlated (see Figure S1). Therefore, instead of the standard Hopfield attractor neural networks, we will implement the projection method neural networks \cite{Kanter1987Associative}.

The correlation between cell fate $\mu$ and $\nu$ is given by
\begin{equation}
 A^{\mu\nu}=\frac{1}{N}\displaystyle\sum_{i=1}^N \xi_i^\mu \xi_i^\nu
 \end{equation}

Now the inferred correlation-based, TF interaction matrix is
 \begin{equation}
 J_{ij}=\frac{1}{N}\displaystyle\sum_{\mu=1}^{p}\displaystyle\sum_{\nu=1}^{p}  \xi_i^\mu (A^{-1})^{\mu\nu} \xi_j^\nu
 \end{equation}
  
 Then the landscape can be rewritten as
\beq
 H&=& -\frac{1}{2}\displaystyle\sum_{i=1}^N \displaystyle\sum_{j \neq i}^N S_i J_{ij} S_j = -\frac{1}{2N}\displaystyle\sum_{i=1}^N \displaystyle\sum_{j \neq i}^N \displaystyle\sum_{\mu=1}^{p}\displaystyle\sum_{\nu=1}^{p}    S_i \xi_i^\mu (A^{-1})^{\mu\nu} \xi_j^\nu S_j \\
 &=& -\frac{N}{2}\displaystyle\sum_{\mu=1}^p m^\mu a^\mu \label{proj_order}
\eeq 
 where in equation \ref{proj_order} we have introduced the projection order parameter $a^\mu$ which is the orthogonal projection of a spin vector onto the subspace spanned by the stable cell fates
 \be
 a^\mu=\displaystyle\sum_{\nu=1}^p (A^{-1})^{\mu\nu}m^\nu = \displaystyle\sum_{\nu=1}^p\displaystyle\sum_{i=1}^N (A^{-1})^{\mu\nu}\xi^\nu_i S_i
 \ee
 
 A simple geometric picture illustrates that $H$ makes each cell fate a global minimum of the landscape.  An arbitrary vector can be rewritten in terms of its projection in the cell fate subspace and its orthogonal component $\delta S_i$,
 \begin{equation}
 S_i = \displaystyle\sum_{\mu=1}^p a^\mu \xi^\mu_i + \delta S_i
 \end{equation}
Then, the distance of an arbitrary vector $\textbf{S}$ to the cell fate subspace is given by $\Delta$,
\begin{equation}
\Delta = \left(\displaystyle\sum_{i=1}^N (\delta S_i)^2 \right)^{1/2}
\end{equation}
which can be rewritten as
\begin{equation}
\frac{\Delta^2}{N} = 1 - \displaystyle\sum_{\mu=1}^p a^\mu m^\mu
\end{equation}
This allows us to rewrite the stabilizing term of the landscape as
\begin{equation}
H = -\frac{N}{2}+\frac{1}{2}\Delta^2
\end{equation}

This provides a very clear interpretation of the landscape as the global distance of an arbitrary vector $\textbf{S}$ to the natural cell fate subspace \cite{Kanter1987Associative}. 

Again, let's examine the signal-to-noise of cell fates in the absence of stochastic update noise. If a state is dynamically stable, one needs $S_i h_i >0$. When the state is a given cell fate (without loss of generality assume cell fate $1$), we have that

\beq
\xi^1_1 h_1 &=& \frac{1}{N} \displaystyle\sum_{j\neq i}^N \displaystyle\sum_{\mu=1}^p \xi^1_1 \xi^\mu_1 \left(A^{-1} \right)^{\mu\nu} \xi^\nu_j \xi^1_j \\
&=&  \displaystyle\sum_{\mu=1}^p \xi^1_1 \xi^\mu_1 \left(A^{-1} \right)^{\mu\nu} A^{\nu1} = 1
\eeq

Therefore, the stability of cell fate $1$ has no noise interference from the other cell fates, and we have that cell fates are stable up to $p/N=1$.

\clearpage{}

\section{Supplementary Figures}

\begin{figure}[h]
\includegraphics[width=15cm]{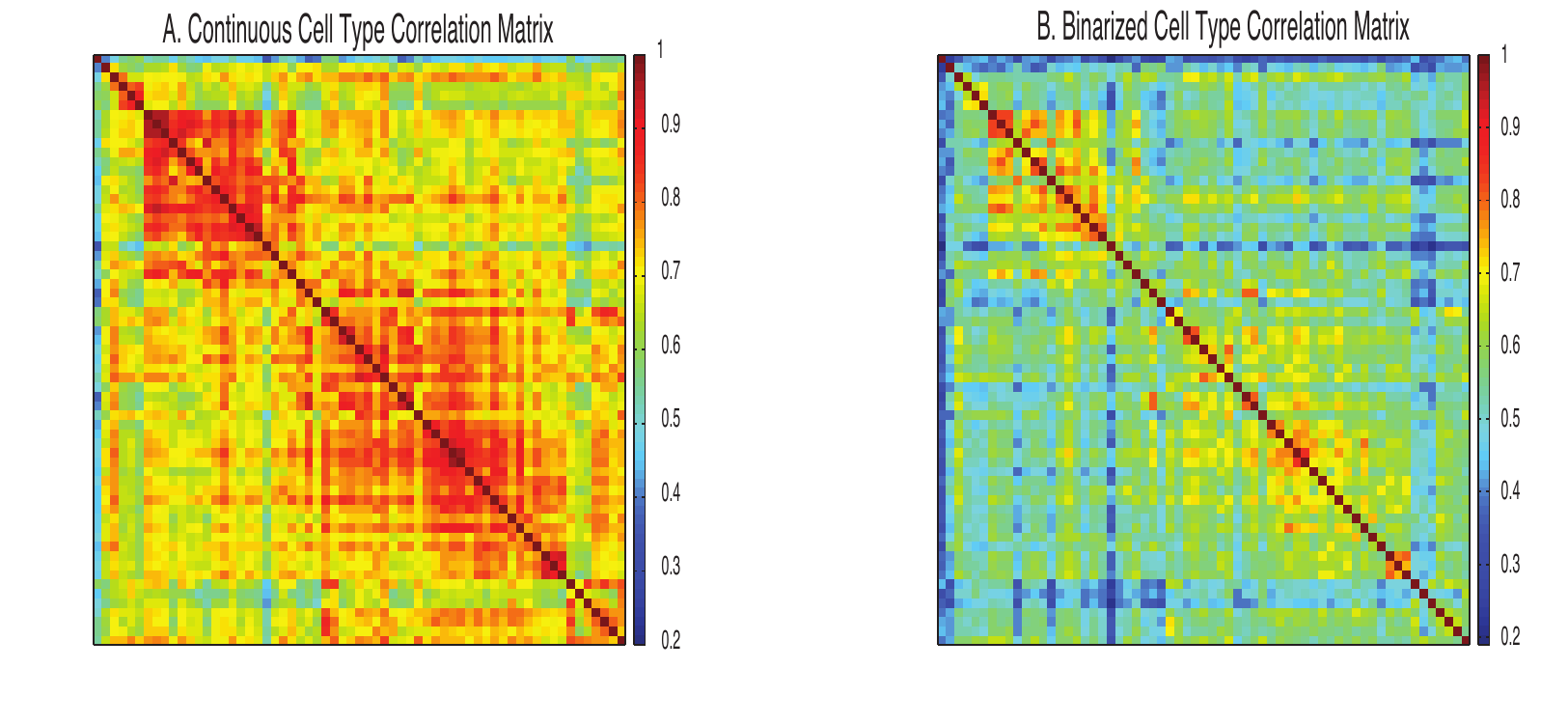}
\caption{\textbf{Cell fate correlation matrices.}}
\textnormal{ (A) Correlation matrix between cell fates for continuous data. (B) Correlation matrix for binarized data.}
\end{figure}

\begin{figure}[h]
\includegraphics[width=15cm]{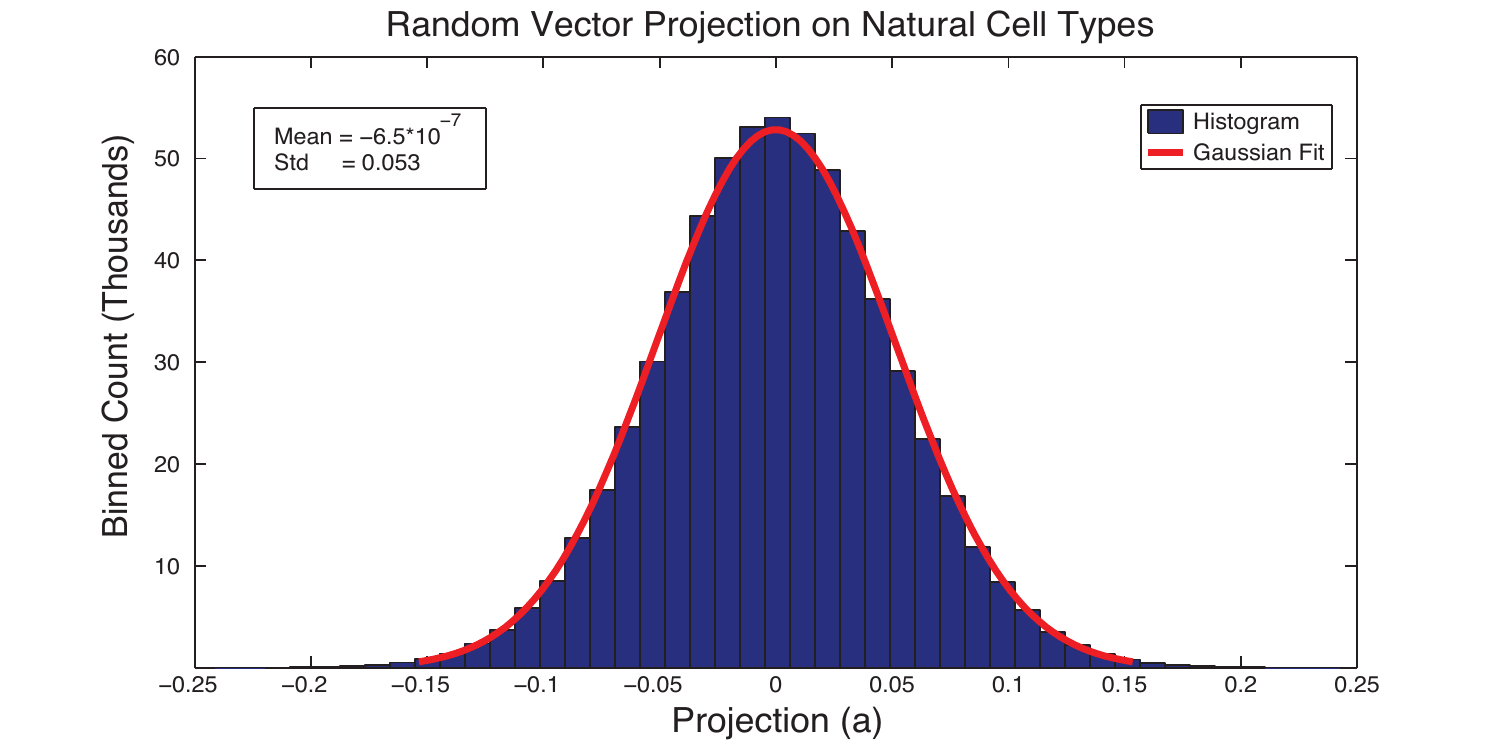}
\caption{\textbf{Projection of a random vector on a given cell fate.}}
\textnormal{   Ten thousand binarized random vectors were created in MATLAB and projected onto the Õcellular sub-spaceÕ. The histogram shows the distribution of the projections. The red line is a Gaussian fit to the histogram. The mean is practically zero while the standard deviation is 0.053..}
\end{figure}

\begin{figure}[h]
\includegraphics[width=15cm]{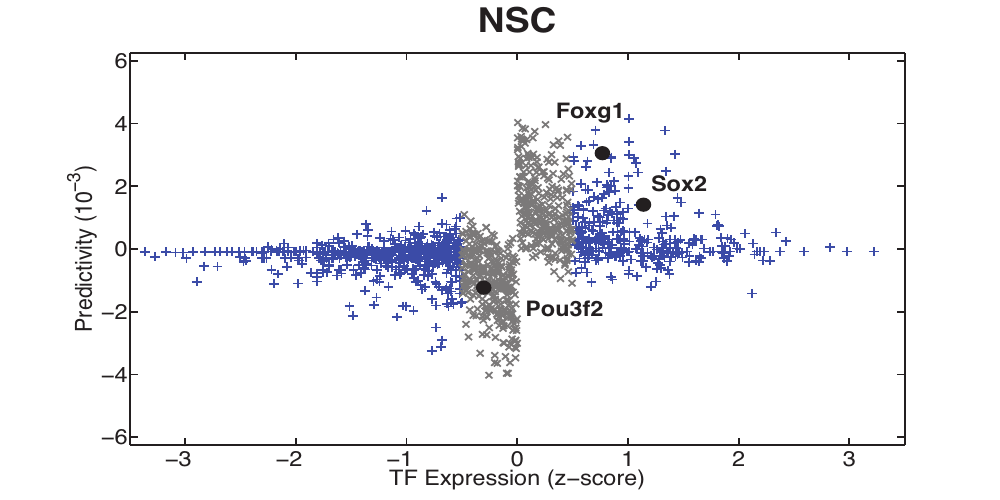}
\caption{\textbf{Predictivity vs Expression for NSC}}
\textnormal{ Same type of plot as Figure 3. Labeled TFs are part of reprogramming protocol to NPC \cite{Lujan2012Direct}. This illustrates that \emph{Foxg1} is predictive for NSC, even though it is not for NPC.}
\end{figure}

\clearpage{}

\section{Table S1}

\begin{table}[h]{\textbf{Table S1. Classifying Top ESC Reprogramming Candidates.}}

\vspace{5 mm}
\begin{tabular}{| c | c | c | c | c | c |}
\hline
TF & Z Score & $\eta$ $(10^{-3})$ & Rank Z$*\eta$ & Classification & Citation or GO Term\\
\hline
\hline
\emph{Pou5f1} (\emph{Oct4})  &  2.77  &  2.45  &  1  &Reprogramming	& \cite{Gonzalez2011Methods}  \\ \hline
\emph{Gm13242}  &  1.59  &  3.98  &  2  &Unknown	&biological process \\ \hline
\emph{Nr0b1}  &  2.44  &  2.59  &  3  &Pluripotency	& \cite{Khalfallah2009Dax-1}\\ \hline
\emph{Nanog}  &  2.30  &  2.65  &  4  &Reprogramming	& \cite{Gonzalez2011Methods} \\ \hline
\emph{Zfp42}  &  2.04  &  2.74  &  5 &Pluripotency	& \cite{Takahashi2006Induction} \\ \hline
\emph{Hsf2bp}  &  1.42  &  3.49  &  6  &Unknown	&biological process  \\ \hline
\emph{Esrrb}  &  1.74  &  2.49  &  7  &Reprogramming	& \cite{Gonzalez2011Methods}  \\ \hline
\emph{Zscan4f}  &  1.01  &  3.86  &  8  &Reprogramming	& \cite{Jiang2013Zscan4}  \\ \hline
\emph{Klf4}  &  1.04  &  3.25  &  9  &Reprogramming	& \cite{Gonzalez2011Methods}  \\ \hline
\emph{Zfp459}  &  0.83  &  3.98  &  10  &Unknown	&biological process  \\ \hline
\emph{Zscan4c}  &  0.82  &  3.86  &  11  & Pluripotency   &   telomere lengthening \\ \hline
\emph{Zic3}  &  1.17  &  2.65  &  12  &Pluripotency	& \cite{Lim2007Zic3}  \\ \hline
\emph{Zfp936}  &  1.15 &  2.66  &  13  &Unknown	&biological process\  \\ \hline
\emph{Zfp229}  &  0.76  &  3.84  &  14  &Unknown	&biological process \\ \hline
\emph{Zfp600 } &  0.71  &  3.98  &  15  &Unknown	&biological process  \\ \hline
\emph{Zfp640}  &  1.10  &  2.55  &  16  & Differentiation   & skeletal system morphogenesis \\ \hline
\emph{Gm10324}  &  1.09  &  2.55  &  17  &Unknown	&biological process    \\ \hline
\emph{Zscan10}  &  1.04  &  2.65  &  18 &Pluripotency 	& \cite{Yu2009Zfp206}\ \\ \hline
\emph{Utf1}  &  2.03  &  1.30  &  19 &Reprogramming	& \cite{Gonzalez2011Methods}  \\ \hline
\emph{2610305D13Rik}  &  1.02  &  2.45  &  20  &Unknown	&biological process  \\ \hline
\emph{Tfcp2l1}  &  1.26  &  1.90  &  21  &  Pluripotency  & \cite{Ye2013Embryonic} \\ \hline
\emph{Klf8}  &  0.58  &  4.12  &  22  &  Differentiation  & \cite{Lee2012Kruppel-Like}  \\ \hline
\emph{Epas1}  &  0.70  &  3.18  &  23  & Differentiation   &  erythrocyte differentiation \\ \hline
\emph{Tbx3}  &  1.09  &  2.03  &  24  &Reprogramming	& \cite{Gonzalez2011Methods} \\ \hline
\emph{Tcf15}  &  0.89  &  2.37  &  25  &Differentiation	 & \cite{Davies2013Tcf15}\\ \hline
\end{tabular}
\end{table}

Table has top 50 embryonic stem cell (ESC) reprogramming candidates (as ranked by z-score times predictivity, $\eta_i^\mu$). Classification of each TF is either justified by paper citation or GO Process term. Reprogramming TFs are in a pre-existing reprogramming protocol, pluripotency TFs help maintain the ESC state but are non-essential for reprogramming, differentiation TFs are expressed in ESC but help induce cell fate change \emph{in vivo}, and unknown TFs have no known function.

\clearpage{}

\begin{table}[h]\textbf{Table S1 Continued. Classifying Top ESC Reprogramming Candidates.}
\vspace{5 mm}
\begin{tabular}{| c | c | c | c | c | c |}
\hline
TF & Z Score & $\eta$ $(10^{-3})$ & Rank Z$*\eta$ & Classification & Citation or GO Term\\
\hline
\hline
\emph{Tcfl5}  &  0.82  &  2.56  &  26  &Unknown	&regulation of transcription \\ \hline
\emph{Sall4 } &  1.72  &  1.17  &  27  &Reprogramming	&\cite{Gonzalez2011Methods} \\ \hline
\emph{Zfp553}  &  0.87  &  2.22  &  28  &Unknown	&regulation of transcription  \\ \hline
\emph{Sox2}  &  1.96  &  0.97  &  29  &Reprogramming	& \cite{Gonzalez2011Methods}  \\ \hline
\emph{Grhl3}  &  0.61  &  2.75  &  30  &Differentiation	 &ectoderm development  \\ \hline
\emph{Zbtb10}  &  0.75  &  2.22  &  31  &Unknown	&negative regulation of transcription  \\ \hline
\emph{Mycn}  &  1.90  &  0.85  &  32  &  Differentiation  & lung development  \\ \hline
\emph{Sap30}  &  0.93  &  1.72  &  33  &  Differentiation  & skeletal muscle cell differentiation \\ \hline
\emph{Zbtb8a}  &  0.83  &  1.88  &  34  &Unknown	&regulation of transcription \\ \hline
\emph{Klf5}  &  1.23  &  1.25  &  35  &Differentiation    & skeletal muscle cell differentiation  \\ \hline
\emph{Sall1}  &  1.30  &  1.18  &  36  &Differentiation	& neural tube development  \\ \hline
\emph{AA987161}  &  0.60  &  2.36  &  37  &Unknown    &biological process  \\ \hline
\emph{Klf9}  &  0.70  &  1.96  &  38  & Differentiation   & embryo implantation \\ \hline
\emph{Myc}  &  0.73  &  1.86  &  39  &Reprogramming	& \cite{Gonzalez2011Methods}   \\ \hline
\emph{Rarg}  &  0.87  &  1.54  &  40 &Differentiation & bone morphogenesis \\ \hline
\emph{Tead2} & 1.03 & 1.15 & 41 &Differentiation  & lateral mesoderm development \\ \hline
\emph{Dnmt3b}  &  1.33  &  0.88  &  42  &Pluripotency	& genetic imprinting \\ \hline
\emph{Nr5a2}  &  0.67  &  1.75  &  43  &Reprogramming	&\cite{Gonzalez2011Methods}  \\ \hline
\emph{Nr1d2}  &  0.74  &  1.53  &  44  &Differentiation	&regulation of skeletal muscle cell differentiation  \\ \hline
\emph{Cbx7}  &  1.14  &  0.99  &  45  & Differentiation   & chromatin modification \\ \hline
\emph{Bnip3}  &  1.40  &  0.77  &  46  & Differentiation   & brown fat cell differentiation\\ \hline
\emph{Rbpms}  &  1.63  &  0.64  &  47  & Unknown   & transcription, DNA-templated\\ \hline
\emph{Zfp7}  &  0.91  &  1.15  &  48  & Unknown   & regulation of transcription, DNA-templated\\ \hline
\emph{Lin28a}  &  0.78  &  1.31  &  49  &Reprogramming	& \cite{Gonzalez2011Methods}   \\ \hline
\emph{Zfp423}  &  0.55  &  1.79  &  50  &Differentiation	& cell differentiation   \\ \hline
\end{tabular}
\end{table}

\clearpage{}

\section{Table S2}
\begin{table}{\textbf{Table S2. Examining Yamanaka Factors in Detail}}

\vspace{5mm}

 \begin{tabular}{| c | c | c | c | c  | c | c |}
 \hline
 TF & (A) Exp. & (B) Diff. Exp.  & (C) Specificity & & (D) Predictivity  & (E) Exp*Pred  \\
 \hline
 \hline

\emph{Pou5f1} (\emph{Oct4}) & 2 & 1 & 1.6\%	& & 70 & 1\\
\hline
\emph{Sox2}  & 22 & 11 & 0.0\%	&  &201 & 29\\
\hline
\emph{Klf4} & 122 & 124 & 22.2\%	&&  28 & 9\\
\hline
\emph{Myc} & 213 & 1183 & 66.7\%	&&  106 & 39\\
\hline
 \end{tabular}
 \end{table}
 
Here we reexamine the Yamanaka transcription factors (TFs) in light of our model. When the Yamanaka results were first published, \emph{Klf4} and \emph{Myc} were counterintuitive factors \cite{Jaenisch2012Nuclear}. \emph{Myc} was quickly shown to enhance the efficiency of reprogramming but was dispensable \cite{Wernig2008c-Myc}. \emph{Klf4} remained a surprise, but this table demonstrates the power of predictivity by establishing the importance of \emph{Klf4}. The columns (A),(B), and (C) are data about TFs available to Yamanaka, while (D) and (E) are data from our model. Unless otherwise stated, the numbers represent rank order (out of $1337$) relative to the other TFs. To understand the importance of rank order, the original Yamanaka experiment used 24 TFs while most later studies test around 10 TFs at once. (A) Exp. is TF expression rank in embryonic stem cells (ESC). (B) Diff. Exp. is the differential expression rank between ESC and mouse embryonic fibroblasts (MEF), the starting cell fates in the Yamanaka protocol. (C) Specificity is the percentage of cell fates (out of our $63$) which have expression at the same or higher level as the ESC. (D) Predictivity is the novel measure of TF importance generated by our model. (E) Exp*Pred is the rank of the product of expression and predictivity of highly expressed TFs and is an attempt to find a single quantity signifying reprogramming potential. The data available to Yamanaka illustrates that \emph{Pou5f1} (\emph{Oct4}) and \emph{Sox2} were natural choices. \emph{Myc} is an oncogene that enhances proliferation but was found to be non-essential for reprogramming  \cite{Wernig2008c-Myc}, so we will ignore it. The power of predictivity is illustrated by examining \emph{Klf4} which is not highly expressed (A), differentially expressed (B), or specific (C). However, it is very predictive of ESC (D) and is a top choice when examining Exp*Pred (E). Note that \emph{Klf4} illustrates that predictivity is not exactly the same as specificity. While \emph{Klf4} is expressed in many cell fates, since predictivity takes into account correlations between cell fate expression patterns, predictivity can filter out the uncorrelated expression pattern and highlight the importance of \emph{Klf4} for ESC.

\bibliographystyle{plos2009}